%% file: Datta_SciKG_arXiv.tex
\documentclass[onecolumn]{DDLabJournalPaper}

\usepackage{longtable}
\input{Preamble/DDLabIdentity}
\citationauthors{D. Datta}          

\makeatletter
\patchcmd{\DDJ@titleblock}{\LARGE\bfseries}{\Large\bfseries}{}{}
\patchcmd{\DDJ@titleblock}{{\normalsize\DDJ@renderauthors\par}}%
                          {{\large\DDJ@renderauthors\par}}{}{}
\makeatother

\newcommand{\Determined}{\textbf{Determined:}\hspace{0.35em}}
\newcommand{\NotDetermined}{\textbf{Not yet determined:}\hspace{0.35em}}

\usetikzlibrary{decorations.pathmorphing}

\tikzset{
  stnode/.style={ddnode, font=\footnotesize, text width=27mm, minimum height=11mm},
  stwide/.style={stnode, text width=36mm},
  stdec/.style={dddecision, font=\footnotesize, text width=24mm, minimum height=11mm},
  stlab/.style={ddedgelabel, font=\footnotesize},
}

\manuscripttype{Research Article}
\venue{Preprint}
\preprintid{}                       
\manuscriptversion{}

\title{From Autonomous Knowledge Generation to Generalizable Property
       Prediction: A Precision-First Pipeline for Electro-Optic Materials
       Discovery}
\shorttitle{Precision-first knowledge generation for electro-optic discovery}

\DDAuthor{\textbf{Dibakar Datta}}{}
\citationyear{2026}

\DDAffiliation{}{Dibakar Datta Laboratory (DDLab)\\
  New Jersey Institute of Technology (NJIT), Newark, NJ 07103, USA}

\correspondence{\href{mailto:ddlab@njit.edu}{ddlab@njit.edu}}
\date{\today}

\paperabstract{%
Electro-optic materials, whose refractive index changes linearly with an
applied electric field (the Pockels effect), are the hardware that moves
information between electronic and optical signals. They set the speed and
energy of optical modulators and are a key component of the
microwave-to-optical links that a networked quantum computer will need,
often at cryogenic temperature. Designing better ones means optimising a
material's composition together with its operating conditions, such as
temperature, strain, and bias field. No database of electro-optic tensors
exists for a machine-learning model to learn from: general repositories
such as the Materials Project and the Open Quantum Materials Database do not
hold them. We describe a pipeline that builds such a database from the
literature and test whether it supports prediction for materials held out of
training. The pipeline follows one rule, \emph{no value is better than a
wrong value}. It has five stages: physics-aware acquisition, layout-aware
extraction with a large language model (LLM) and deterministic table
parsers, a three-tier verification gate that accepts a value only when it is
positively verified, cross-paper consensus using the median and the median
absolute deviation (MAD), and a noise-robust surrogate model audited by
leave-one-material-out (LOMO) testing. Physics enters the gate directly:
the $\gamma_{ij}$ notation is mapped to $r_{ij}$, Pockels components that
crystal symmetry forbids in cubic zinc-blende crystals are rejected,
symmetry-equivalent labels are merged by point group, and every value
carries a provenance tag. In one production run over 49 publications, the
gate turned 816 candidate values into 414 accepted, 215 held for review, and
187 rejected, each with a typed reason. Against a hand-certified gold
standard of 303 values from 16 papers, the run reaches a precision of 0.51
and a recall of 0.48 when every field of a record must match, so the gate
reduces error but does not remove it. Across four database snapshots, the
number of held-out materials for which the model beat a predict-the-mean
baseline rose from 1 of 4 to 9 of 13 as the database grew from 10 to 22
materials. The snapshots were not designed as a controlled comparison, so we
report this as a trend. The result supports the premise that a
precision-oriented, physics-checked literature database can carry a forward
model beyond the materials it was trained on, and it identifies what is
still missing before that model can be inverted for design.}

\highlights{%
\begin{itemize}[leftmargin=1.2em, itemsep=0.15em, topsep=0.1em]
  \item The first literature-derived, physics-checked database of
        electro-optic (Pockels) tensors built for machine learning, a property
        absent from general materials databases.
  \item A five-stage pipeline (acquire, extract, verify, reconcile, model) in
        which withholding a value is a legitimate output at every stage.
  \item Physics in the verification gate: $\gamma\to r$ symbol recovery,
        zinc-blende symmetry rejection, point-group label merging, and
        provenance tags.
  \item Measured accuracy against a 303-value gold standard (precision 0.51,
        recall 0.48 at full-record match), reported with the gate's own
        tallies.
  \item Leave-one-material-out prediction that improved from 1 of 4 to 9 of 13
        held-out materials as the number of distinct materials grew.
\end{itemize}}

\plainsummary{%
Electro-optic crystals turn electrical signals into changes of light, and
future quantum computers will need them to send information from ultra-cold
processors over optical fibre. Finding better ones with artificial
intelligence requires data that do not yet exist in any database. We built a
system that reads scientific papers, keeps a value only when it passes
physics-based checks, and assembles the survivors into a database. We measure
how often it is still wrong, and we show that a model trained on its output
predicts materials it never saw better as the database covers more kinds of
materials.}

\begin{document}
\maketitle

\section{Introduction}
\label{sec:intro}

\subsection{Why electro-optic materials matter}
\label{sec:why-eo}

An electro-optic material changes its refractive index when an electric
field is applied to it. In the linear (Pockels) effect, the change is
proportional to the field and is described by a third-rank tensor
\citep{nye1985physical, yariv1984optical, boyd2020nonlinear},
\begin{equation}\label{eq:pockels}
  \Delta\!\left(\frac{1}{n^{2}}\right)_{\!i} \;=\; \sum_{j=1}^{3} r_{ij}\,E_{j},
  \qquad i = 1,\dots,6,
\end{equation}
where $(1/n^{2})_i$ are the six independent components of the optical
indicatrix in contracted (Voigt) notation, $E_j$ is the applied field along
crystal axis $j$, and $r_{ij}$ are the Pockels coefficients, usually quoted in
pm/V. Which of the 18 components $r_{ij}$ may be non-zero is fixed by the
crystal's point group, and only crystals without a centre of inversion have
any. A phase modulator of length $L$ and electrode gap $d$ needs a half-wave
voltage
\begin{equation}\label{eq:vpi}
  V_{\pi} \;=\; \frac{\lambda\, d}{n^{3}\, r_{\mathrm{eff}}\, \Gamma\, L},
\end{equation}
where $\lambda$ is the optical wavelength, $n$ the refractive index,
$r_{\mathrm{eff}}$ the effective coefficient for the chosen field and
polarisation geometry, and $\Gamma$ the overlap of the optical and electrical
fields. A larger $n^{3}r_{\mathrm{eff}}$ means a lower drive voltage, a smaller
device, and less energy per bit, which is why lithium niobate and, more
recently, barium titanate on silicon have become the workhorses of integrated
photonics \citep{wang2018integrated, zhu2021integrated, zhang2021integrated,
abel2019large}.
\Cref{fig:eo-physics}(b) shows the effect in a modulator.

\input{Figures/fig_eo_physics}

\paragraph{Electro-optic, optical, and nonlinear-optical materials}
These three terms are often used loosely, and the distinction matters for what
the database stores. \emph{Optical materials} is the broad class of materials
used to transmit, guide, or shape light, characterised by refractive index,
absorption, and dispersion. \emph{Nonlinear-optical} materials respond to the
optical field itself in a way that is not linear in that field; second-order
($\chi^{(2)}$) processes such as second-harmonic generation and parametric
down-conversion mix optical frequencies with each other. \emph{Electro-optic}
materials are those whose optical properties are controlled by an external,
low-frequency or static electric field. The Pockels effect is the limit of the
same $\chi^{(2)}$ response in which one of the two fields is at, or near, zero
frequency, and the quadratic (Kerr) effect is its third-order counterpart
\citep{boyd2020nonlinear}. A crystal can be an excellent nonlinear-optical
material and still be a poor modulator, so the database records Pockels and
effective electro-optic coefficients as their own property family and treats
nonlinear-optical susceptibilities as out of scope.

\paragraph{The nervous system of a quantum computer}
If quantum materials are the brain of a quantum computer, the media in which
qubits are made and controlled \citep{basov2017towards}, electro-optic
materials are its nervous system. Superconducting qubits operate at millikelvin
temperatures and are controlled and read out by microwave signals, whereas
low-loss communication over distance is optical. Linking processors into a
network, or even moving their signals out of a dilution refrigerator without
a bundle of heat-carrying coaxial cables, needs a coherent converter between
microwave and optical photons \citep{lauk2020perspectives, han2021microwave} (\cref{fig:eo-physics}a).
Cavity electro-optic transducers do this through the Pockels effect
\citep{holzgrafe2020cavity, sahu2022quantum}, and cryogenic electro-optic
links already read out superconducting circuits optically
\citep{youssefi2021cryogenic}. In each case the conversion efficiency, the
added noise, and the heat load are set by the Pockels coefficient that the
material retains at a few kelvin, and that coefficient is not the
room-temperature value printed in handbooks. In barium titanate it drops by
roughly a factor of two on cooling but stays far above lithium niobate
\citep{eltes2020integrated}; in strontium titanate, a quantum paraelectric,
thin films engineered for low temperature reach very high values
\citep{ulrich2025engineering}; and epitaxial strain changes it again
\citep{paoletta2021pockels, suceava2026colossal}.

\paragraph{The design problem: a material and its knobs}
The Pockels response is therefore not a single number per compound. It depends
on the composition and phase, on the crystal or film form and the substrate,
and on external knobs: temperature, mechanical strain or clamping, a bias field
that can induce a large linear response from a quadratic one (as in
potassium tantalate niobate \citep{imai2007crystal}), and the optical
wavelength and modulation frequency. Designing an electro-optic material means
searching jointly over the material configuration $\vect{c}$ and the operating
conditions $\vect{k}$,
\begin{equation}\label{eq:designproblem}
  (\vect{c}^{\star},\vect{k}^{\star}) \;=\;
  \argmax_{\vect{c}\in\Set{C},\;\vect{k}\in\Set{K}}\;
  n^{3}(\vect{c},\vect{k})\; r_{\mathrm{eff}}(\vect{c},\vect{k}),
  \qquad \vect{k}=(T,\ \varepsilon,\ E_{\mathrm{bias}},\ \lambda,\ f),
\end{equation}
for example the configuration with the largest Pockels coefficient at
\SI{4}{K} under a strain the substrate can supply, where $T$ is temperature,
$\varepsilon$ strain, $E_{\mathrm{bias}}$ the bias field, $\lambda$ the
wavelength, and $f$ the modulation frequency. \Cref{fig:design-space} sketches
this joint space.

\input{Figures/fig_design_space}

\paragraph{Why there is little AI for electro-optics}
Machine learning has accelerated discovery where large, consistent property
databases exist. The Materials Project \citep{jain2013materials}, the Open
Quantum Materials Database (OQMD) \citep{saal2013oqmd}, AFLOW
\citep{curtarolo2012aflow}, and the Inorganic Crystal Structure Database
(ICSD) \citep{belsky2002icsd} standardise structures, formation energies, and
electronic descriptors, and the Materials Project also holds computed
elastic, piezoelectric, and dielectric tensors \citep{dejong2015elastic,
dejong2015piezo, petousis2017high}. None of them holds electro-optic tensors,
and none records the temperature, strain, field, and wavelength at which a
functional property was measured. Those data exist only in papers, reported
under incompatible conventions ($r_{ij}$, $\gamma_{ij}$, effective
coefficients, clamped and unclamped values) and printed alongside dielectric,
elastic, and piezoelectric constants that are easily confused with them. Without
a database, \cref{eq:designproblem} cannot be posed to a model at all. Building
that database, and showing that it supports prediction beyond the materials it
contains, is the purpose of this paper.

\subsection{The data bottleneck}

The knowledge that governs functional behaviour remains largely in
unstructured publications: the synthesis route, crystallographic orientation,
phase, measurement method, and the conditions of measurement
\citep{tshitoyan2019unsupervised, olivetti2020data,
kononova2021opportunities}. A property value has scientific meaning only when
it is bound to its material, its tensor component, and its conditions.
Automated literature mining has become practical. Rule-based toolkits such as
ChemDataExtractor \citep{swain2016chemdataextractor} have produced databases of
magnetic phase-transition temperatures \citep{court2018auto} and battery
materials \citep{huang2020database}, and text mining has recovered synthesis
routes \citep{kim2017materials}. Large language models now extract structured
records from scientific text with accuracy approaching that of expert curators
\citep{polak2024extracting, dagdelen2024structured}. The same models can also
attach a correct number to the wrong label, material, or condition, or produce
a value that is not in the source \citep{ji2023hallucination}. For a tensor
property, where one wrong component can corrupt every model trained on it,
that failure matters more than coverage.

In earlier work we introduced an autonomous scientific knowledge generation
framework that turns literature into an AI-ready knowledge base through
ontology-guided acquisition, hybrid extraction, and semantic harmonisation
\citep{datta2026framework}. That work set out the architecture on a small
corpus and left two questions open. First, how accurate are the values the
system emits, and can a verification step stop the physically impossible ones?
Second, does the resulting database support prediction for materials held out
of training, which is the property any generative design step depends on? This
paper answers both for inorganic electro-optic materials. A pipeline that
extracts their tensors correctly has to reason about symmetry, notation, and
measurement context, which are the same capabilities needed for batteries,
catalysis, and quantum materials.

\begin{KeyIdea}
Checking physics before a value enters the database is part of building the
predictive model, not only a matter of tidy data. A model can predict a
material it has never seen only by transferring what it learned from other
materials. One physically impossible value in training, such as a Pockels
component that the crystal's symmetry forbids, shifts the learned relation for
every neighbouring composition, and no later step can tell it from a real
measurement. Rejecting such values at the point of acceptance is therefore a
modelling decision.
\end{KeyIdea}

The contributions of this paper are:
\begin{itemize}[leftmargin=1.2em, itemsep=0.2em]
  \item A five-stage pipeline (physics-aware acquisition, layout-aware
        extraction, a three-tier verification gate, cross-paper consensus, and
        noise-robust modelling) in which each stage can refuse to pass a value
        forward, with a flowchart and an algorithm for each stage
        (\cref{sec:framework,sec:methods}).
  \item Physics embedded in verification: reversible symbol recovery that maps
        $\gamma_{ij}$ to $r_{ij}$, rejection of Pockels components forbidden in
        cubic zinc-blende crystals, point-group merging of symmetry-equivalent
        labels, and provenance tags that separate measured, computed, and
        device-effective values (\cref{sec:gate}).
  \item A hand-certified gold standard drawn from 16 papers
        (\cref{sec:gold}) and the measured accuracy of the pipeline against
        it, reported with the gate's own tallies, so that precision is a
        number rather than a design intention (\cref{sec:accuracy}).
  \item Evidence that leave-one-material-out prediction improves as the number
        of distinct materials grows, from 1 of 4 to 9 of 13 held-out materials
        across four database snapshots (\cref{sec:lomo}), and a concrete path
        from this forward model to inverse design over materials and operating
        conditions (\cref{sec:discussion}).
\end{itemize}

A companion paper \citep{datta2026erroneous} analyses the errors that survive
this pipeline, with a taxonomy, a layered detection framework, and an audit of
accepted values against reference tensors. The two papers share one production
run and use the same accuracy figures.

\section{Framework}
\label{sec:framework}

\subsection{The pipeline}

The pipeline converts a set of publications into a set of canonical property
records through five stages (\cref{fig:pipeline}). Stage~0 finds and ranks
candidate papers before any are read; Stages~1 to~4 extract, verify,
reconcile, and model. Each of Stages~1 to~4 may withhold a value rather than
emit an uncertain one. \Cref{sec:methods} describes each stage with its own
flowchart.

\begin{DDWideFigure}[tb]
  \centering
  \begin{tikzpicture}[node distance=6mm and 10mm,
      every node/.append style={font=\small}]
    \node[ddevidence, text width=25mm, minimum height=16mm] (acq)
      {\textbf{0.\ Acquire}\\\footnotesize query, rank, download};
    \node[ddmethod, text width=29mm, minimum height=16mm, right=of acq] (ext)
      {\textbf{1.\ Extract}\\\footnotesize text, tables,\\material, conditions};
    \node[dddecision, text width=29mm, minimum height=16mm, right=of ext] (gate)
      {\textbf{2.\ Verify}\\\footnotesize gate: physics, context, symbol};
    \node[ddstructure, text width=29mm, minimum height=16mm, right=of gate] (cons)
      {\textbf{3.\ Reconcile}\\\footnotesize cross-paper consensus};
    \node[ddresult, text width=29mm, minimum height=16mm, below=16mm of cons] (model)
      {\textbf{4.\ Model}\\\footnotesize robust surrogate,\\LOMO test};
    \node[ddcontrol, text width=29mm, minimum height=16mm, below=16mm of gate] (hold)
      {\textbf{Review / reject}\\\footnotesize typed reasons, never deleted};
    \draw[ddflow] (acq) -- (ext);
    \draw[ddflow] (ext) -- (gate);
    \draw[ddgate] (gate) -- node[ddedgelabel, above, font=\footnotesize]{accept} (cons);
    \draw[ddflow] (cons) -- (model);
    \draw[ddreject] (gate) -- node[ddedgelabel, right, font=\footnotesize]{hold or reject} (hold);
    \draw[ddreturn] (model) -- node[ddedgelabel, above, font=\footnotesize, align=center]
      {flagged\\records} (hold);
    \draw[ddreturn] (ext.south) |- node[ddedgelabel, pos=0.75, below, font=\footnotesize]
      {papers held at detection} (hold.west);
  \end{tikzpicture}
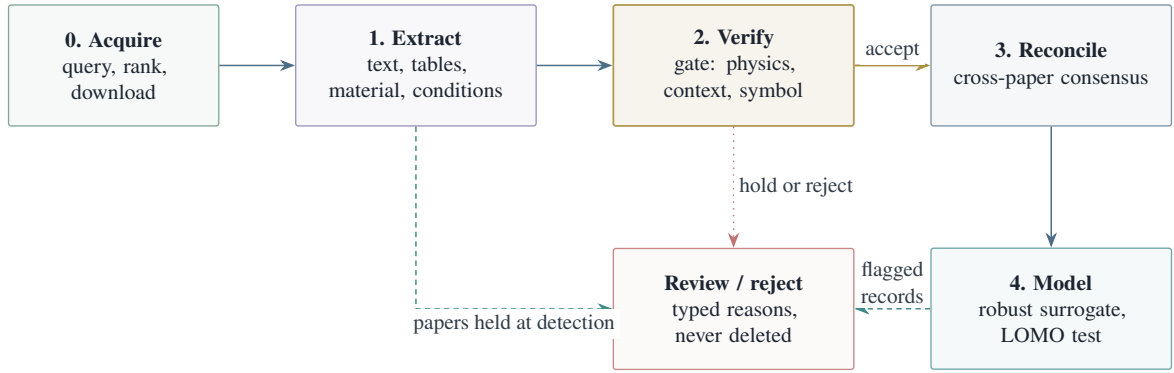
  \caption{The five-stage pipeline. Stage~0 assembles the corpus; extraction
  recovers candidate records with their material and measurement context; the
  gate accepts only positively verified records and routes the rest to review
  or rejection with a typed reason; consensus reconciles records across papers
  and weights them; the surrogate is trained on the weighted records and tested
  by leave-one-material-out (LOMO) prediction. Dashed edges return papers and
  records to the review list, where they are kept with their reasons.}
  \label{fig:pipeline}
\end{DDWideFigure}

\begin{KeyTakeaway}
Every stage can say ``this value cannot be verified'' and keep the value, with
its reason, outside the database. In a conventional extract-then-clean
pipeline every extracted value enters the database and errors are searched for
afterwards, by which point a wrong tensor component looks exactly like a real
one.
\end{KeyTakeaway}

\subsection{Records, scope, and the acceptance rule}
\label{sec:scope}

The unit the pipeline handles is a \emph{record}, not a bare number. A record is
the tuple
\begin{equation}\label{eq:record}
  \rho = (m,\ \ell,\ v,\ u,\ \vect{c},\ \pi,\ s),
\end{equation}
where $m$ is the material (for example LiNbO$_3$), $\ell$ the canonical tensor
label (for example $r_{33}$), $v$ the numerical value, $u$ its unit, $\vect{c}$
the measurement conditions (temperature, strain or mechanical boundary,
wavelength, frequency, form, substrate), $\pi$ the provenance (experiment,
first-principles calculation, fitted set, derived, or unspecified), and $s$
the source (paper and page). The database targets inorganic crystalline
electro-optic, ferroelectric, and related materials; organic and hybrid
systems, whose conventions and symmetry warrant a separate schema, are routed
out at detection. Within scope, one acceptance rule is applied.

\begin{KeyEquation}{Acceptance rule}[applied to every record]
\begin{equation}\label{eq:accept}
  \mathrm{accept}(\rho) \iff
  \underbrace{P(\rho)}_{\text{physics}} \wedge
  \underbrace{C(\rho)}_{\text{context}} \wedge
  \underbrace{S(\rho)}_{\text{symbol}} ,
\end{equation}
\end{KeyEquation}

\noindent
where $P(\rho)$ holds if the value is physically admissible for the material's
symmetry and property family, $C(\rho)$ if the record carries the context that
makes it interpretable and is the paper's own result, and $S(\rho)$ if its
symbol has been resolved to a canonical label. A record that fails any clause
goes to review or is rejected, never into the database. Recall is recovered
later, by consensus and by a modelling stage that tolerates residual noise,
rather than by lowering the bar for acceptance.

\begin{EvidenceNote}
\Determined{} a physically impossible value (a symmetry-forbidden Pockels
component, an out-of-scope quantity) can be rejected deterministically without a
person in the loop. \NotDetermined{} that every accepted record is correct. The
gold-standard measurement in \cref{sec:accuracy} shows that it is not, which is
why consensus weighting and the modelling-stage audit exist.
\end{EvidenceNote}

\section{Methods}
\label{sec:methods}

\subsection{Stage 0: physics-aware literature acquisition}
\label{sec:stage0}

Stage~0 (\cref{fig:stage0}) issues bibliographic queries to Crossref, arXiv,
and PubMed so that coverage does not depend on a single index. Each query pairs
a target material or an electro-optic term with a context term. Returned
records are de-duplicated by digital object identifier (DOI), or by normalised
title when no DOI exists, against a persistent SQLite catalogue. A relevance
ranker then scores each record from its title, abstract, and keywords.

\begin{figure}[tb]
  \centering
  \begin{tikzpicture}[node distance=6mm and 5mm]
    \node[stnode, ddevidence] (q) {Query plan\\{\scriptsize materials $\times$ EO terms}};
    \node[stnode, ddmethod, right=of q] (r) {Retrieve\\{\scriptsize Crossref, arXiv, PubMed}};
    \node[stnode, ddmethod, right=of r] (d) {De-duplicate\\{\scriptsize DOI or normalised title}};
    \node[stnode, ddmethod, right=of d] (s) {Score relevance\\{\scriptsize $R(p)$, \cref{eq:relevance}}};
    \node[stdec, below=9mm of s] (t) {$R(p)\ge\tau$?};
    \node[stnode, ddmethod, left=of t] (dl) {Download\\{\scriptsize open-access PDF}};
    \node[stnode, ddmethod, left=of dl] (v) {Validate file\\{\scriptsize PDF header, SHA-256}};
    \node[stnode, ddresult, left=of v] (cat) {Catalogue\\{\scriptsize metadata, score, embedding}};
    \node[stnode, ddcontrol, below=7mm of t] (skip) {Skipped, logged\\{\scriptsize score kept}};
    \draw[ddflow] (q) -- (r);
    \draw[ddflow] (r) -- (d);
    \draw[ddflow] (d) -- (s);
    \draw[ddflow] (s) -- (t);
    \draw[ddgate] (t) -- node[stlab, above]{yes} (dl);
    \draw[ddflow] (dl) -- (v);
    \draw[ddflow] (v) -- (cat);
    \draw[ddreject] (t) -- node[stlab, right]{no} (skip);
  \end{tikzpicture}
  \caption{Stage~0, acquisition. Queries go to three bibliographic services;
  records are de-duplicated against the catalogue, scored by
  \cref{eq:relevance}, and downloaded only above the threshold $\tau$. A file is
  catalogued only if it is a valid PDF, and its SHA-256 hash detects
  byte-identical duplicates.}
  \label{fig:stage0}
\end{figure}

The score rewards direct electro-optic evidence and penalises vocabulary that
shares surface words but not the target physics. For a paper $p$, let
$n_k(p)$ be the number of matches in its title, abstract, and keywords to
pattern group $k$: direct electro-optic terms (``electro-optic'', ``Pockels'',
explicit coefficients such as $r_{33}$), target crystals, ferroelectric
physics, thin-film growth, quantum-device context (cryogenic, transduction,
qubit), and computational electro-optics. Then
\begin{equation}\label{eq:relevance}
  \begin{aligned}
  R(p) &= \operatorname{clip}_{[0,1]}\!\bigl(\varrho(p)/30\bigr),\\
  \varrho(p) &= \sum_{k} w_k \min\bigl(n_k(p),\,c_k\bigr)
     + \sum_{(k,k')} b_{kk'}\,\mathbf{1}[n_k>0 \wedge n_{k'}>0]
     - \eta(p)\,\pi_{\mathrm{nlo}} \min(n_{\mathrm{nlo}},5)
     - \pi_{\mathrm{out}} \min(n_{\mathrm{out}},4),
  \end{aligned}
\end{equation}
where the group weights are $w_k = 5.0$ (direct electro-optic), 4.0 (target
material), 2.0 (ferroelectric), 1.7 (thin film), 2.5 (quantum device), and
1.2 (computational), each count is capped at $c_k = 4$ (5 for ferroelectric)
so that one repeated word cannot dominate, and the bonuses
$b_{kk'}$ reward the co-occurrences that identify a target paper: electro-optic
term with target material ($+5$), material with ferroelectric physics ($+3$),
material with thin film ($+2$), electro-optic term with quantum device ($+3$),
and electro-optic term with computation ($+2$). $n_{\mathrm{nlo}}$ counts
generic nonlinear-optics terms (second-harmonic generation, parametric
oscillation and down-conversion, squeezing) with penalty
$\pi_{\mathrm{nlo}} = 2.5$, scaled by $\eta(p) = 0.25$ when the paper also has
a direct electro-optic term and a target material, 0.6 when it has one of the
two, and 1 otherwise; $n_{\mathrm{out}}$ counts out-of-scope terms with
penalty $\pi_{\mathrm{out}}=4.0$. Here $\varrho(p)$ is the raw score and $\mathbf{1}[\cdot]$ equals 1 when its condition holds and 0 otherwise. Two gates override the sum: a record with no
direct electro-optic, material, ferroelectric, or quantum-device signal scores
zero, and a record whose only signal is generic nonlinear optics is capped at a
raw score of 2. The divisor 30 is the raw score of a typical strong
electro-optic materials paper (18 to 35 in practice), so $R$ reads as a
fraction of that. Records with $R(p) \ge \tau = 0.15$ are downloaded from
open-access sources; the ranked list shown to the user uses 0.20. Each stored
paper keeps its metadata, score, keywords, and a text embedding for semantic
search.

\begin{ModelingNote}[How the corpus in this paper was assembled]
The 49-publication corpus analysed here (\cref{app:corpus}) was collected
before Stage~0 was coupled to the extraction stages; Stage~0 runs as a
separate program whose output is copied into the extraction input folder.
Stage~0 is described because it is the route by which the corpus is being
scaled. Its precision is the subject of \cref{sec:outlook}: in a first pass,
most of the papers it retrieved did not report a Pockels coefficient.
\end{ModelingNote}

\subsection{Stage 1: layout-aware extraction}
\label{sec:extract}

Each publication is parsed page by page into text that preserves reading order
across two-column layouts, with tables recovered as structured rows and page
numbers kept for provenance (\cref{fig:stage1}). Text normalisation repairs
ligatures and character encodings without splitting words or formulas.

\begin{figure}[tb]
  \centering
  \begin{tikzpicture}[node distance=6mm and 5mm]
    \node[stnode, ddevidence] (pdf) {PDF\\{\scriptsize from Stage 0}};
    \node[stnode, ddmethod, right=of pdf] (parse) {Parse layout\\{\scriptsize reading order, tables, pages}};
    \node[stnode, ddmethod, right=of parse] (det) {Detect material\\{\scriptsize \cref{alg:detect}}};
    \node[stdec, right=of det] (conf) {In scope and confident?};
    \node[stnode, ddcontrol, below=8mm of conf] (held) {Paper held\\{\scriptsize manual assignment}};
    \node[stnode, ddmethod, below=8mm of det] (ctx) {Paper context\\{\scriptsize form, substrate, $T$, $f$}};
    \node[stnode, ddmethod, below left=8mm and -4mm of ctx] (tab) {Table parsers\\{\scriptsize Pockels matrix, constant tables}};
    \node[stnode, ddmethod, below right=8mm and -4mm of ctx] (llm) {LLM per page\\{\scriptsize GPT-4o, JSON schema}};
    \node[stnode, ddresult, below=30mm of ctx] (merge) {Merge\\{\scriptsize deterministic row preferred}};
    \node[stnode, ddresult, right=of merge] (cand) {Candidate records\\{\scriptsize to Stage 2}};
    \draw[ddflow] (pdf) -- (parse);
    \draw[ddflow] (parse) -- (det);
    \draw[ddflow] (det) -- (conf);
    \draw[ddreject] (conf) -- node[stlab, right]{no} (held);
    \draw[ddgate] (conf.south west) -- node[stlab, above left]{yes} (ctx.north east);
    \draw[ddflow] (ctx) -- (tab);
    \draw[ddflow] (ctx) -- (llm);
    \draw[ddflow] (tab) -- (merge);
    \draw[ddflow] (llm) -- (merge);
    \draw[ddflow] (merge) -- (cand);
  \end{tikzpicture}
  \caption{Stage~1, extraction. The subject material is identified first,
  because a wrong material label corrupts every value a paper contributes. A
  paper whose material cannot be identified with confidence is held and none of
  its values are extracted. Deterministic parsers and the LLM then read the
  paper in parallel, and their outputs are merged.}
  \label{fig:stage1}
\end{figure}

\paragraph{Subject-material detection}
Before any value is extracted, the paper's subject material is identified from
a curated vocabulary $\Set{M}$ of 47 inorganic materials (\cref{alg:detect}).
Each material $m \in \Set{M}$ is listed with its \emph{surface forms} $F(m)$:
the ways it is written in papers, such as formulas (LiNbO$_3$), names
(lithium niobate), and abbreviations (LN). Detection is conservative, because
a mislabelled material silently corrupts every value a paper contributes. Let
$T$ be the full text of the paper, $T_0$ its early text (the first 2500
characters of page~1, which contain the title, authors, and abstract), and
$T_{\mathrm{t}}$ the title zone (the first 600 characters of $T_0$). With
$n(f,X)$ the number of whole-word matches of form $f$ in text $X$, the score of
material $m$ is
\begin{equation}\label{eq:detect}
  s_m = \sum_{f\in F(m)\setminus A}\Bigl[n(f,T) + \alpha\, n(f,T_0) +
        \beta\, n(f,T_{\mathrm{t}})\Bigr]
        \;+\; 0.3\,\min\Bigl(1,\ \sum_{f\in F(m)\cap A} n(f,T)\Bigr),
  \qquad \alpha=3,\ \beta=40 .
\end{equation}
Because $T_{\mathrm{t}}\subset T_0\subset T$, a mention in the title counts
$1+\alpha+\beta = 44$ times, a mention in the abstract 4 times, and a mention in
the body once. The weights encode where a paper names its own subject. A paper
on a new crystal routinely compares it with lithium niobate or barium titanate
dozens of times in the body, but names its own material in the title, so
$\beta=40$ makes one title mention outweigh about forty body mentions of a
reference crystal. The abstract weight $\alpha=3$ gives early prominence extra
credit without letting a single comparison in the abstract (``larger than
LiNbO$_3$'') decide the result. The set $A$ holds 26 ambiguous acronyms (such
as KDP, BTO, STO, LN, PZT, KTN, and SiC) that also have unrelated meanings.
They earn at most 0.3, which breaks ties between materials that already have a
strong score but cannot create a winner on its own. The values of $\alpha$,
$\beta$, and 0.3 were set by inspecting detection errors on the corpus, not by
optimisation.

Two guards protect the result. A \emph{gated abbreviation} $(m,a,F_c)$ credits
an abbreviation $a$ that a paper coins for its own crystal (for example
``KTN'' for KTa$_{1-x}$Nb$_x$O$_3$) only when one of the confirming formulas
$F_c$ also occurs in $T$, so a coined abbreviation cannot hijack a paper about
the parent compound. The \emph{confidence guard} requires the winner to occur in
the early text $T_0$. A winner that never appears there is almost always a
cited comparison in a paper whose true subject is outside the vocabulary (an
organic or halide-perovskite paper, for example). Such a paper is \emph{held}:
it is placed on a list for manual assignment, and none of its values are
extracted until a person assigns the material. A paper in which no material
dominates ($s_{m^\star} < 1.5\,s_{m'}$ for the runner-up $m'$) is tagged as a
comparison paper but is still processed, because the confidence guard has
already established that its winner is named early.

\begin{DDAlgorithm}[tb]{Subject-material detection with a confidence guard}{alg:detect}
\Require full text $T$; early text $T_0$ (first 2500 characters of page 1);
  title zone $T_{\mathrm{t}}$ (first 600 characters of $T_0$); vocabulary
  $\Set{M}$ with surface forms $F(m)$ for each material $m$; ambiguous acronyms
  $A$; gated abbreviations $G$ of triples $(m, a, F_c)$
\Ensure subject material $m^{\star}$, a comparison tag, and a confidence flag
\ForAll{$m \in \Set{M}$}
  \State $s_m \gets \sum_{f\in F(m)\setminus A} [\,n(f,T) + 3\,n(f,T_0) + 40\,n(f,T_{\mathrm{t}})\,]$
     \Comment{strong forms, \cref{eq:detect}}
  \State $e_m \gets \sum_{f\in F(m)\setminus A} n(f,T_0)$
     \Comment{early hits, for the guard}
  \State $u_m \gets \sum_{f\in F(m)\cap A} n(f,T)$
     \Comment{ambiguous forms: weak hits}
\EndFor
\ForAll{$(m, a, F_c) \in G$}
  \If{some $f \in F_c$ occurs in $T$}
    \State $s_m \gets s_m + n(a,T) + 3\,n(a,T_0) + 40\,n(a,T_{\mathrm{t}})$;\quad
           $e_m \gets e_m + n(a,T_0)$ \Comment{gated credit}
  \EndIf
\EndFor
\State $s_m \gets s_m + 0.3\min(1,u_m)$ for every $m$ with $s_m>0$
   \Comment{tie-breaker only}
\If{$s_m = 0$ for all $m$}
  \State \Return none, \Term{held} \Comment{no in-scope material}
\EndIf
\State $m^{\star} \gets \argmax_m s_m$;\quad $m' \gets$ the runner-up
\State $\mathit{comparison} \gets [\,s_{m^{\star}} < 1.5\, s_{m'}\,]$
   \Comment{tag only; the paper is processed}
\If{$e_{m^{\star}} = 0$}
  \State \Return $m^{\star}$, $\mathit{comparison}$, \Term{unconfident}
     \Comment{paper held for manual assignment}
\EndIf
\State \Return $m^{\star}$, $\mathit{comparison}$, \Term{confident}
\end{DDAlgorithm}

Word-boundary anchoring stops short acronyms (``STO'', ``LN'', ``SiC'') from
matching inside other words. A separate detector records paper-level context
(form, substrate, temperature, frequency), which is later attached to each
record.

\paragraph{Language-model extraction}
Values are read from each page by a general-purpose LLM (GPT-4o, OpenAI) at
temperature zero in JSON (JavaScript Object Notation) mode, under a schema that
names the subject material, the allowed property families and tensor labels,
and the condition fields to fill. In the production run the cost was under ten
US cents per paper.

\paragraph{Deterministic table parsers}
Two structures that free-text reading handles poorly are read by deterministic
parsers. The first is the bracketed Pockels matrix, in which a full tensor is
printed as a $6\times3$ array of bare numbers whose indices are implied by
position (as in \citep{paoletta2021pockels}); the parser assigns each entry its
$(i,j)$ label from its row and column. The second is a constant table of
electro-optic, elastic, or refractive values with explicit subscripts. When a
deterministic row and an LLM row report the same tensor and value, or the same
tensor under the same condition, the deterministic row is kept. In the
production run, 62 of the 414 accepted values came from these parsers and 352
from the LLM.

\begin{ModelingNote}
Adding a vision pass over rendered page images was net-negative. Against the
gold standard of \cref{sec:gold}, it raised recall from 0.37 to 0.43 but lowered
precision from 0.46 to 0.38, because the model over-generates from figures and
schematics that carry no extractable coefficient \citep{datta2026erroneous}.
Consistent with the acceptance rule, extraction runs on a single text pass, and
structured content is handled by the deterministic parsers rather than by a
vision model.
\end{ModelingNote}

\subsection{Stage 2: the three-tier verification gate}
\label{sec:gate}

Every candidate record is assigned to one of three tiers (accepted, review, or
rejected) by the conjunction of \cref{eq:accept} (\cref{fig:stage2,alg:gate}).
The gate is positive: a record is accepted only if it matches a recognised
label pattern for its property family, carries a unit consistent with that
family, and passes the physics and symbol clauses below. A record that is not
verified is kept with a typed reason, so that recall can be recovered by a
later decision.

\begin{figure}[tb]
  \centering
  \begin{tikzpicture}[node distance=5mm and 9mm]
    \node[stwide, ddevidence] (in) {Candidate record $\rho$ from Stage 1};
    \node[stwide, ddmethod, below=of in] (sym) {$S$: recover symbol\\{\scriptsize $\gamma\to r$, superscripts, merge by point group}};
    \node[stdec, text width=36mm, below=of sym] (lab) {Label recognised; family, prefix, unit consistent?};
    \node[stdec, text width=36mm, below=of lab] (phys) {$P$: in physical range and allowed by symmetry?};
    \node[stdec, text width=36mm, below=of phys] (ctx) {$C$: paper's own result, verbatim, keyed, plausible?};
    \node[stwide, ddresult, below=of ctx] (acc) {\textbf{Accepted}\\{\scriptsize provenance and condition completeness tagged}};
    \node[ddnode, ddcontrol, font=\footnotesize, text width=58mm, right=of lab] (r1)
      {\textbf{Rejected}: corrupted or index-only label, equation number\\
       \textbf{Review}: missing indices, unit or family mismatch};
    \node[ddnode, ddcontrol, font=\footnotesize, text width=58mm, right=of phys] (r2)
      {\textbf{Rejected}: out of range, forbidden component (\cref{eq:zb}), out-of-scope family, OCR-corrupted};
    \node[ddnode, ddcontrol, font=\footnotesize, text width=58mm, right=of ctx] (r3)
      {\textbf{Review}: cited or comparison value, series value without key, not verbatim, implausible magnitude};
    \draw[ddflow] (in) -- (sym);
    \draw[ddflow] (sym) -- (lab);
    \draw[ddgate] (lab) -- node[stlab, right]{yes} (phys);
    \draw[ddgate] (phys) -- node[stlab, right]{yes} (ctx);
    \draw[ddgate] (ctx) -- node[stlab, right]{yes} (acc);
    \draw[ddreject] (lab) -- node[stlab, above]{no} (r1);
    \draw[ddreject] (phys) -- node[stlab, above]{no} (r2);
    \draw[ddreject] (ctx) -- node[stlab, above]{no} (r3);
  \end{tikzpicture}
  \caption{Stage~2, the verification gate. A record passes the symbol clause
  $S$, a label and unit check, the physics clause $P$, and the context clause
  $C$ in order, and is accepted only if it passes all of them. A record that is
  impossible (a corrupted label, a symmetry-forbidden component, an
  out-of-range value) is rejected; one that is merely unverified is held for
  review. Both keep their reasons.}
  \label{fig:stage2}
\end{figure}

\paragraph{Symbol recovery ($S$)}
Notation is normalised before it is judged. Underscores and run-together
boundary superscripts are removed reversibly
($p\_122\to p_{122}$, $\varepsilon^{S}_{11}$ written as \Code{eS11}), without
contracting indices. The classical symbol $\gamma_{ij}$ for the linear
electro-optic coefficient is mapped to $r_{ij}$, and combined coefficients
$\gamma_c$ to an effective coefficient $r_{\mathrm{eff}}$. This rule recovered a
five-component tensor from a hydrothermal KTiOPO$_4$ (KTP) paper
\citep{zhou2023hydrothermal} that reported $\gamma_{33}$, $\gamma_{23}$,
$\gamma_{13}$, $\gamma_{42}$, and $\gamma_{51}$; the recovered
$\gamma_{33}=\SI{36.3}{pm/V}$ matched a value the same run had already
accepted as $r_{33}=\SI{36.3}{pm/V}$.

\paragraph{Symmetry admissibility ($P$)}
Crystal symmetry fixes which components of $r_{ij}$ in \cref{eq:pockels} may be
non-zero \citep{nye1985physical}. For cubic zinc-blende crystals (point group
$\bar{4}3m$), namely GaAs, InP, GaP, ZnTe, ZnSe, CdTe and their alloys, the
tensor has a single independent component:
\begin{equation}\label{eq:zb}
  \bigl[r_{ij}\bigr]_{\bar{4}3m} =
  \begin{pmatrix}
    0 & 0 & 0\\ 0 & 0 & 0\\ 0 & 0 & 0\\
    r_{41} & 0 & 0\\ 0 & r_{41} & 0\\ 0 & 0 & r_{41}
  \end{pmatrix},
  \qquad r_{52}=r_{63}=r_{41}.
\end{equation}
Any other $r_{ij}$ reported for such a material is rejected as an extraction
error. Out-of-scope families (phonon frequencies, band gaps, optical loss) are
rejected by the same guard (\cref{lst:gate}). Each property family also has a
physical range outside which a value is rejected: \SIrange{0.05}{6000}{pm/V}
for electro-optic coefficients, 0.05 to \SI{3000}{pC/N} for piezoelectric
strain coefficients, 1 to \SI{2000}{GPa} for elastic constants, 1 to 200\,000
for relative permittivity, and 1 to 6 for refractive index. The ranges are
deliberately wide; they catch unit and decimal errors, not unusual materials.

\begin{lstlisting}[language=Python, caption={The physics clauses of the gate as
implemented. Symmetry rejection is currently implemented for the zinc-blende
class only.}, label={lst:gate}]
ZB43M = {"GaAs","InP","GaP","ZnTe","ZnSe","CdTe","ZnxCd1-xTe","Cd1-xMgxTe"}
ZB_ALLOWED = {"r41", "r52", "r63", "r_eff"}         # zinc-blende (-43m)
OUT_OF_SCOPE = {"phonon_frequency", "band_gap", "optical_loss", "raman"}

def admissible(material, family, label):
    if family in OUT_OF_SCOPE:                       # not a database property
        return False
    if material in ZB43M and family == "electro_optic" \
       and label not in ZB_ALLOWED:                  # symmetry-forbidden component
        return False
    return True
\end{lstlisting}

Symmetry also governs how a correctly read value is labelled. Two labels that
the point group forces to be equal are merged into one canonical label, a step
we call \emph{folding}. Writing $\phi_g$ for the folding map of point group
$g$,
\begin{equation}\label{eq:fold}
  \phi_{3m}: \{r_{23}\mapsto r_{13},\ r_{42}\mapsto r_{51},\
             r_{12},r_{61}\mapsto r_{22}\},\qquad
  \phi_{4mm}: \{r_{23}\mapsto r_{13}\},\qquad
  \phi_{\bar{4}3m}: \{r_{52},r_{63}\mapsto r_{41}\},
\end{equation}
applied to the trigonal $3m$ crystals of the corpus (LiNbO$_3$, LiTaO$_3$,
BiFeO$_3$, K$_3$Nb$_3$B$_2$O$_{12}$), the tetragonal $4mm$ crystals
(BaTiO$_3$, PIN--PMN--PT, SrTiO$_3$), and zinc-blende. For $4mm$ crystals
$r_{42}$ and $r_{51}$ are kept as separate labels to match the reference
tables the gold standard follows; for orthorhombic $mm2$ KNbO$_3$ no
components are equivalent and none are folded. On the electro-optic part of
the gold standard, folding raised recall from 0.593 to 0.658 while precision
was essentially unchanged (0.478 to 0.475). Folding relabels; it does not
reject. A component that is forbidden in a non-cubic class is not yet rejected
(\cref{sec:limits}).

\paragraph{Unit consistency}
A value is admitted for a property family only if the unit reported with it is
dimensionally consistent with that family. The rule was added after a
permittivity of several thousand, read from a dense multi-property table whose
column header sat several rows away, was assigned to an elastic-constant slot.
An elastic value whose source text carries no elastic unit is now held.

\paragraph{Provenance and context ($C$)}
Each accepted record is tagged as experiment, first-principles calculation,
fitted self-consistent set, derived relation, or unspecified. Device-level
coefficients are labelled $r_{\mathrm{eff}}$ and marked \Term{effective}, apart
from intrinsic tensor components. Values the source cites from earlier work are
held, so that a well-known constant is not counted again in every paper that
quotes it. A value from a series (for example, one of several temperatures)
without the key that identifies its member is held, as is a value that does not
occur verbatim in the source text. Conditions are scored as complete, partial,
or missing, and a bulk single crystal is marked as the zero-strain reference so
that ``unstrained'' is distinguishable from ``strain not reported''.

\begin{DDAlgorithm}[tb]{The three-tier verification gate (rules applied in order)}{alg:gate}
\Require candidate record $\rho=(m,\ell,v,u,\vect{c},\pi,s)$ with its raw
  source snippet $x$; point group $g(m)$; family ranges $[a_F,b_F]$
\Ensure tier $\in\{$\Term{accepted}, \Term{review}, \Term{rejected}$\}$ and reasons
\State $\ell \gets \phi_{g(m)}(\mathrm{normalise}(\ell))$
   \Comment{$S$: $\gamma\to r$, superscripts, \cref{eq:fold}}
\If{$\ell$ is corrupted, index-only, or an equation or reference number}
  \State \Return \Term{rejected}
\EndIf
\If{the family of $\rho$ is out of scope \OR $x$ is OCR-corrupted}
  \State \Return \Term{rejected}
\EndIf
\If{$m$ is zinc-blende \AND $\ell\notin\{r_{41}, r_{\mathrm{eff}}\}$}
  \State \Return \Term{rejected} \Comment{$P$: \cref{eq:zb}}
\EndIf
\If{$v \notin [a_F, b_F]$}
  \State \Return \Term{rejected} \Comment{$P$: physical range}
\EndIf
\If{$\ell$ lacks indices \OR family, prefix, and unit disagree}
  \State add reason; tier $\gets$ \Term{review}
\EndIf
\If{$v$ is cited \OR a literature comparison \OR from a series without a key}
  \State add reason; tier $\gets$ \Term{review} \Comment{$C$}
\EndIf
\If{$v$ is not verbatim in $x$ \OR an elastic value lacks an elastic unit}
  \State add reason; tier $\gets$ \Term{review}
\EndIf
\If{an electro-optic bulk value exceeds \SI{250}{pm/V} \OR a permittivity exceeds $10^{4}$}
  \State add reason; tier $\gets$ \Term{review} \Comment{plausibility}
\EndIf
\If{tier is unset}
  \State tag provenance $\pi$ and condition completeness; \Return \Term{accepted}
\EndIf
\State \Return \Term{review} with all reasons
\end{DDAlgorithm}

\begin{ModelingNote}
The gate is deliberately strict at the point of acceptance. It will hold a
correct value whose label it cannot positively recognise, such as an unusual
symbol or a missing index. A false rejection costs recall, which later stages
can partly recover; a false acceptance costs correctness, which nothing
downstream can recover reliably.
\end{ModelingNote}

\subsection{Stage 3: cross-paper consensus}
\label{sec:consensus}

Several papers often report the same quantity, for example $r_{33}$ of
LiNbO$_3$ at room temperature. \emph{Consensus} here means the agreement among
the independent reports of one quantity: a value that agrees with the other
papers is more likely to be a correct reading than one that stands alone or
disagrees. Stage~3 (\cref{fig:stage3,alg:consensus}) measures that agreement
and turns it into a confidence weight $w$ for each record.

\begin{figure}[tb]
  \centering
  \begin{tikzpicture}[node distance=6mm and 5mm]
    \node[stnode, ddevidence] (in) {Accepted records\\{\scriptsize from Stage 2}};
    \node[stnode, ddmethod, right=of in] (grp) {Group by key $\kappa(\rho)$\\{\scriptsize \cref{eq:groupkey}}};
    \node[stdec, right=of grp] (n3) {Group size $n\ge3$?};
    \node[stnode, ddmethod, right=of n3] (mad) {Median, MAD, $z_i$\\{\scriptsize \cref{eq:mad}}};
    \node[stdec, below=9mm of mad] (out) {$|z_i|>3.5$ and $\delta_i>0.15$?};
    \node[stnode, ddresult, left=of out] (inl) {Inlier\\{\scriptsize $w=\min(1,0.5+0.1n)\,a$}};
    \node[stnode, ddresult, below=9mm of n3] (single) {Single source\\{\scriptsize $w=0.5$}};
    \node[stnode, ddcontrol, below=7mm of out] (rev) {Consensus review\\{\scriptsize $w\le0.2$, kept}};
    \draw[ddflow] (in) -- (grp);
    \draw[ddflow] (grp) -- (n3);
    \draw[ddgate] (n3) -- node[stlab, above]{yes} (mad);
    \draw[ddflow] (n3) -- node[stlab, right]{no} (single);
    \draw[ddflow] (mad) -- (out);
    \draw[ddgate] (out) -- node[stlab, above]{no} (inl);
    \draw[ddreject] (out) -- node[stlab, right]{yes} (rev);
  \end{tikzpicture}
  \caption{Stage~3, consensus. Accepted records that describe the same
  quantity are grouped; in groups of three or more, a robust score identifies
  values that disagree with the rest. Every record leaves this stage with a
  confidence weight; none is deleted.}
  \label{fig:stage3}
\end{figure}

Records are grouped by the key
\begin{equation}\label{eq:groupkey}
  \kappa(\rho) = \bigl(m,\ F,\ \ell,\ \operatorname{sign}(v),\ \tau(T)\bigr),
\end{equation}
where $F$ is the property family and $\tau(T)$ a coarse temperature regime:
cryogenic below \SI{150}{K}, room temperature up to \SI{400}{K}, and
\SI{100}{K} bins above. The mechanical boundary condition (clamped or
unclamped) can be added to the key; this split is off by default because many
papers do not state it. Within a group of $n\ge3$ reports, agreement is
assessed with the modified $z$-score of Iglewicz and Hoaglin
\citep{iglewicz1993outliers, leys2013detecting}, built on the median and the
MAD, which are insensitive to the outliers they must flag:

\begin{KeyEquation}{Robust consensus score}[median--MAD]
\begin{equation}\label{eq:mad}
  z_i = \frac{0.6745\,(v_i - \tilde{v})}{\mathrm{MAD}},\qquad
  \mathrm{MAD} = \operatorname{median}_j |v_j - \tilde{v}|,\qquad
  \delta_i = \frac{|v_i-\tilde v|}{|\tilde v|},
\end{equation}
\end{KeyEquation}

\noindent
where $\tilde{v}$ is the group median and the factor 0.6745 makes $z_i$
comparable to a standard normal score. A record is an outlier when
$|z_i|>3.5$ and its relative deviation $\delta_i$ exceeds 0.15; when the MAD is
zero (most reports identical), the condition $\delta_i>0.15$ alone applies. The
second condition stops a group of nearly identical values from flagging a
report that differs only in the last digit. The confidence weight is
\begin{equation}\label{eq:consweight}
  w_i =
  \begin{cases}
    \min(1,\ 0.5+0.1\,n)\times a & \text{inlier in a group of } n\ge3,\\
    0.5 & \text{single source } (n<3),\\
    0.2 & \text{unit conflict within the group},\\
    0.15 & \text{consensus outlier},\\
    0.1 & \text{out of physical bounds, or no usable value},
  \end{cases}
\end{equation}
where $a$ is the fraction of inliers in the group, so a fully agreeing group
of three gives $w=0.8$ and of five or more gives $w=1$.

The low weights ($w\le0.2$) are chosen to keep a disputed record in the
database without letting it steer the model. A record that disagrees with its
group may be a transcription error, or it may be a correct but unusual
measurement, for example a different domain state or a different wavelength
that the group key does not separate. Frequency alone cannot tell the two
apart, so deleting outliers would remove exactly the measurements that might
matter most. With $w\le0.2$, a disputed record contributes at most a fifth of
the pull of a verified one in the weighted loss of Stage~4, and less than half
that of a single-source record ($w=0.5$), whose only weakness is that nobody
else has measured it. The ordering, not the exact values, is what matters:
agreed $>$ single source $>$ disputed $>$ impossible. Outliers and unit
conflicts are also placed on a consensus review list.

\begin{DDAlgorithm}[tb]{Cross-paper consensus weighting}{alg:consensus}
\Require accepted records $\{\rho_i\}$; group key $\kappa$ (\cref{eq:groupkey})
\Ensure confidence weight $w_i$ for every record; a consensus review list
\ForAll{groups $\Set{G} = \{\rho_i : \kappa(\rho_i) = k\}$}
  \State $n \gets |\Set{G}|$
  \If{$n < 3$}
    \State $w_i \gets 0.5$ for all $\rho_i\in\Set{G}$ \Comment{single source}
    \State \textbf{continue}
  \EndIf
  \State $\tilde v \gets \operatorname{median}(v)$;\quad
         $\mathrm{MAD} \gets \operatorname{median}|v-\tilde v|$
  \ForAll{$\rho_i \in \Set{G}$}
    \State $\delta_i \gets |v_i-\tilde v|/|\tilde v|$
    \If{$\mathrm{MAD}>0$}
      \State $o_i \gets [\,|0.6745(v_i-\tilde v)/\mathrm{MAD}| > 3.5 \wedge \delta_i > 0.15\,]$
    \Else
      \State $o_i \gets [\,\delta_i > 0.15\,]$
    \EndIf
  \EndFor
  \State $a \gets$ fraction of $\rho_i\in\Set{G}$ with $o_i$ false
  \ForAll{$\rho_i \in \Set{G}$}
    \If{$o_i$}
      \State $w_i \gets 0.15$; add $\rho_i$ to the review list
    \Else
      \State $w_i \gets \min(1,\ 0.5+0.1n)\cdot a$
    \EndIf
  \EndFor
\EndFor
\end{DDAlgorithm}

\subsection{Stage 4: noise-robust modelling and the generalization test}
\label{sec:model}

The consensus database feeds a forward surrogate model $f_\theta$ that
predicts $y=\log_{10}|r|$ for electro-optic records (\cref{fig:stage4}).
The inputs $\vect{x}$ are of three kinds: composition features
(fraction-weighted mean and range of atomic number, atomic mass,
electronegativity, and atomic radius, plus the number of elements), so that the
model can score a formula absent from training; condition features (temperature
in kelvin, strain in percent, wavelength, clamped or unclamped boundary, film or
bulk, effective or intrinsic, computed or measured); and the canonical tensor
label.

\begin{figure}[tb]
  \centering
  \begin{tikzpicture}[node distance=6mm and 5mm]
    \node[stnode, ddevidence] (db) {Weighted records\\{\scriptsize from Stage 3}};
    \node[stnode, ddmethod, right=of db] (feat) {Features $\vect{x}$\\{\scriptsize composition, conditions, label}};
    \node[stnode, ddmethod, right=of feat] (wt) {Row weights $\omega_i$\\{\scriptsize \cref{eq:weight}}};
    \node[stnode, ddmethod, right=of wt] (fit) {Fit $f_\theta$\\{\scriptsize gradient boosting, Huber loss}};
    \node[stnode, ddmethod, below=9mm of fit] (oof) {5-fold audit\\{\scriptsize residual $z_i$, \cref{eq:oof}}};
    \node[stnode, ddmethod, below=9mm of wt] (lomo) {LOMO loop\\{\scriptsize hold out each material}};
    \node[stnode, ddresult, left=of lomo] (dom) {Verdicts, domain $\Set{D}_\tau$\\{\scriptsize \cref{eq:lomo}}};
    \node[stnode, ddcontrol, below=7mm of oof] (rev) {Review queue\\{\scriptsize $|z_i|>2.5$}};
    \draw[ddflow] (db) -- (feat);
    \draw[ddflow] (feat) -- (wt);
    \draw[ddflow] (wt) -- (fit);
    \draw[ddflow] (fit) -- (oof);
    \draw[ddflow] (fit) -- (lomo);
    \draw[ddflow] (lomo) -- (dom);
    \draw[ddreturn] (oof) -- (rev);
  \end{tikzpicture}
  \caption{Stage~4, modelling. A robust regressor is trained on weighted
  records. An out-of-fold audit returns records the model finds incompatible
  with their neighbours to review, and the leave-one-material-out (LOMO) loop
  produces a per-material verdict that defines where the model may be trusted.}
  \label{fig:stage4}
\end{figure}

Each row is weighted by its consensus weight and by how complete its
conditions are,
\begin{equation}\label{eq:weight}
  \omega_i = w_i \cdot k(\vect{c}_i), \qquad
  k = 1.0,\ 0.7,\ 0.4 \ \text{for complete, partial, missing conditions},
\end{equation}
where $w_i$ is the consensus weight of \cref{eq:consweight}. The regressor is
gradient boosting \citep{friedman2001greedy} (300 trees, depth 3, learning rate
0.05, subsample 0.8) trained on the weighted Huber loss \citep{huber1964robust}
\begin{equation}\label{eq:huber}
  \Set{L}(\theta) = \sum_i \omega_i\, h_\delta\bigl(y_i - f_\theta(\vect{x}_i)\bigr),\qquad
  h_\delta(e) =
  \begin{cases}
    \tfrac12 e^{2}, & |e|\le\delta,\\
    \delta\,|e| - \tfrac12\delta^{2}, & |e|>\delta,
  \end{cases}
\end{equation}
with $\delta$ set at each boosting step to the 90th percentile of the absolute
residuals. The loss is quadratic for ordinary errors and linear for large ones,
so a residual extraction error pulls the fit far less than it would under
squared error.

The model then audits the labels it was trained on. A five-fold out-of-fold
prediction $\hat y_i^{\,\mathrm{oof}}$ is made for every record, from a model
that did not see it, and the residuals $e_i = y_i - \hat y_i^{\,\mathrm{oof}}$
are standardised robustly,
\begin{equation}\label{eq:oof}
  z_i^{\mathrm{res}} = \frac{0.6745\,(e_i - \tilde e)}{\operatorname{median}_j|e_j-\tilde e|},
\end{equation}
with $\tilde e$ the median residual. Records with $|z_i^{\mathrm{res}}|>2.5$
are listed as incompatible with their chemical neighbours
\citep{northcutt2021confident}. The list is a review queue, not a deletion rule.
An ensemble of eight bootstrap random forests supplies a second signal, the
spread $\sigma_\theta(\vect{x})$ of their predictions, which is large where the
data are thin or contradictory.

The decisive test is leave-one-material-out generalization. Let $\Set{M}_{\mathrm{db}}$
be the set of materials in the database. For a material $m \in \Set{M}_{\mathrm{db}}$, let
$y_m$ be the vector of its electro-optic targets and $\vect{x}_m$ its inputs;
let $f_{\theta\setminus m}$ be the model trained on every record except those of
$m$, and $\bar{y}_{\setminus m}$ the mean target over those same records. All of
$m$ is removed from training, so the model has never seen that material, and
its error is compared with a baseline that predicts the training mean:

\begin{KeyEquation}{Out-of-distribution generalization}[the metric that matters]
\begin{equation}\label{eq:lomo}
  \mathrm{generalizes}(m) \iff
  \mathrm{MAE}\bigl(y_m,\, f_{\theta\setminus m}(\vect{x}_m)\bigr) <
  \mathrm{MAE}\bigl(y_m,\, \bar{y}_{\setminus m}\bigr),
\end{equation}
\end{KeyEquation}

\noindent
where MAE is the mean absolute error. The test is run for every material with
at least three electro-optic records.

\begin{PhysicsInsight}
Random cross-validation can look predictive while the model has only memorised
patterns within each material. LOMO asks whether knowledge of other materials
predicts a material never seen, which is the question an inverse-design step
depends on. The predict-the-mean baseline is a minimum bar; stronger baselines
are discussed in \cref{sec:limits}.
\end{PhysicsInsight}

\subsection{The gold standard}
\label{sec:gold}

A gate reports what it stopped, not what it let through. To measure the
latter we built a gold standard: a set of records read by hand from a fixed
set of papers, independent of any extraction pipeline, against which the
pipeline's output can be scored.

\paragraph{Choice of papers}
Sixteen papers from the corpus were chosen to span the situations an
extractor meets, rather than to be typical (\cref{tab:gold}). They include
the dense reference tensor sets for BaTiO$_3$, KNbO$_3$, and LiNbO$_3$, in
which electro-optic, elastic, piezoelectric, dielectric, elasto-optic, and
refractive constants share one table \citep{zgonik1994dielectric,
zgonik1993materials, jazbinsek2002material}; first-principles studies of
BaTiO$_3$ that print full Pockels matrices as a function of strain or lattice
constant \citep{kim2023nature, paoletta2021pockels}; measurements on bulk
crystals of LiTaO$_3$, PIN--PMN--PT, K$_3$Nb$_3$B$_2$O$_{12}$, and the
zinc-blende semiconductors \citep{casson2004electro, liu2024large, lv2025new,
jeon2007electro, berozashvili1989dispersion}; thin-film and device papers on
BaTiO$_3$, BiFeO$_3$, SrTiO$_3$, and 3C-SiC, where a single effective value
sits among many device figures \citep{suceava2026colossal,
chelladurai2025barium, zhu2016ferroelectric, ulrich2025engineering,
powell2022integrated}; and a review of potassium tantalate niobate
\citep{imai2007crystal}.

\paragraph{Certification}
Each value was read from the paper by hand and recorded with the page, the
verbatim source text, the material, the tensor label as printed and its
canonical form, the unit, the mechanical boundary condition (clamped or
unclamped), the contribution type (electronic, ionic, piezoelectric, or
total), the provenance, and the measurement conditions. Values the paper cites
from other work were excluded, so that the gold standard, like the gate,
contains each paper's own results only. Some papers contribute very few gold
values, and that is deliberate. The potassium tantalate niobate review reports
a Pockels coefficient of about \SI{2200}{pm/V} that is induced from the Kerr
coefficient by a bias field \citep{imai2007crystal}, not an intrinsic linear
coefficient, so it contributes no gold Pockels value; it tests whether a
pipeline wrongly accepts one. The device papers similarly contribute one or two
values each and many tempting numbers that are not coefficients.

\paragraph{Matching}
A predicted record counts as correct only if its value, material, canonical
component, boundary condition, and contribution type all match a gold record
from the same paper; this is the full-record match used throughout. Numeric
precision, which counts a number as correct regardless of its label, is
reported separately because it isolates the errors that put the right number
in the wrong slot.

\begin{table}[tb]
  \caption{The 16 papers of the gold standard, with the material and the kind
  of source each represents.}
  \label{tab:gold}
  \centering
  \small
  \begin{tabular}{L{0.28\linewidth} L{0.44\linewidth} l}
    \toprule
    \DDHeadRow
    \DDHead{Material} & \DDHead{Kind of source} & \DDHead{Reference} \\
    \midrule
    BaTiO$_3$                  & bulk reference tensor set (all families) & \citep{zgonik1994dielectric} \\
    KNbO$_3$                   & bulk reference tensor set (all families) & \citep{zgonik1993materials} \\
    LiNbO$_3$                  & bulk reference tensor set (all families) & \citep{jazbinsek2002material} \\
    BaTiO$_3$                  & first principles, tetragonal, strain     & \citep{kim2023nature} \\
    BaTiO$_3$                  & first principles, Pockels matrices       & \citep{paoletta2021pockels} \\
    BaTiO$_3$                  & strained thin film, cryogenic            & \citep{suceava2026colossal} \\
    BaTiO$_3$                  & thin film and device, MHz to sub-THz     & \citep{chelladurai2025barium} \\
    LiTaO$_3$                  & bulk crystal, near infrared              & \citep{casson2004electro} \\
    PIN--PMN--PT               & single-domain bulk crystal               & \citep{liu2024large} \\
    K$_3$Nb$_3$B$_2$O$_{12}$   & new bulk crystal                         & \citep{lv2025new} \\
    BiFeO$_3$                  & thin-film modulator                      & \citep{zhu2016ferroelectric} \\
    SrTiO$_3$                  & thin film, cryogenic                     & \citep{ulrich2025engineering} \\
    SiC (3C)                   & integrated modulator                     & \citep{powell2022integrated} \\
    CdTe, Zn$_x$Cd$_{1-x}$Te   & bulk crystal, \SI{1550}{nm}              & \citep{jeon2007electro} \\
    GaP                        & bulk crystal, dispersion                 & \citep{berozashvili1989dispersion} \\
    KTa$_{1-x}$Nb$_x$O$_3$     & review; field-induced response (control) & \citep{imai2007crystal} \\
    \bottomrule
  \end{tabular}
\end{table}

\section{Results}
\label{sec:results}

\subsection{What the gate did in a production run}
\label{sec:run}

The most recent single production run (September 2026) processed the 49 PDFs
of \cref{app:corpus}. Three papers were withheld at detection as out of scope
(hybrid or organic perovskites, or a multi-material survey), and 43 papers
yielded at least one candidate value. The 816 candidates were divided into 414
accepted, 215 held for review, and 187 rejected, each held or rejected value
with at least one typed reason (\cref{tab:tiers}). The 414 accepted values come
from 35 papers and 19 materials. They comprise 232 electro-optic values and 182
values of other families: refractive index (49), elastic (33), piezoelectric
strain (30) and stress (11), elasto-optic (27), dielectric (16), optical
dielectric (14), and density (2).

\begin{table}[tb]
  \caption{Typed reasons attached to held and rejected values in the September
  2026 production run (816 candidate values; 414 accepted, 215 held, 187
  rejected). A value can carry several reasons, so columns can sum to more than
  the number of values. Reproduced from the companion analysis
  \citep{datta2026erroneous}.}
  \label{tab:tiers}
  \centering
  \small
  \begin{tabular}{L{0.56\linewidth} r r r}
    \toprule
    \DDHeadRow
    \DDHead{Reason class} & \DDHead{Held} & \DDHead{Rejected} & \DDHead{Total} \\
    \midrule
    Corrupted, missing, or unrecognised tensor label        & 95 & 100 & 195 \\
    Out-of-scope property family (e.g.\ phonon frequencies) &  0 & 137 & 137 \\
    Cited or comparison value, not the paper's own result   & 54 &  32 &  86 \\
    Missing condition key (value from a series)             & 27 &   0 &  27 \\
    Family, label, or unit mismatch                         & 25 &   2 &  27 \\
    Implausible magnitude for the stated context            & 19 &   3 &  22 \\
    Symmetry-forbidden component (zinc-blende rule)         &  0 &  20 &  20 \\
    Optical character recognition (OCR) corruption, or value not verbatim in the source &  5 &  12 &  17 \\
    Not a measurement (equation number, imaginary mode)     &  0 &   4 &   4 \\
    \bottomrule
  \end{tabular}
\end{table}

The provenance of the accepted electro-optic values matters for any model
trained on them. Of the 232, 67 are tagged as experiment, 90 as first-principles
calculation, 10 as part of a fitted self-consistent set, 4 as derived, and 61
as unspecified; 207 are intrinsic tensor components and 25 effective device
coefficients. Only 50 carry complete temperature-and-strain context, 107 partial,
and 75 none.

\subsection{Measured extraction accuracy}
\label{sec:accuracy}

We compared the production output with the certified gold standard of
\cref{sec:gold}, 303 values from 16 papers, using the full-record match.

\begin{KeyFinding}[Measured accuracy]
At the level of individual target values, the production configuration reaches
a precision of 0.51 and a recall of 0.48 against the gold standard. On an
earlier run, numeric precision (the number alone) was 0.74, and nearly half of
the false positives were the right number stored in the wrong slot: the wrong
tensor label, boundary condition, or contribution type \citep{datta2026erroneous}.
\end{KeyFinding}

These figures set the scope of everything that follows. The gate removes whole
classes of error deterministically (\cref{tab:tiers,sec:caught}), but about half
of what it accepts still differs from the gold standard in at least one field.
An audit of the accepted values from three dense legacy tensor papers found that
30 of 112 were defective, and that checks needing no reference data, chiefly
applying symmetry rules to every crystal class rather than only to zinc-blende,
would have caught 21 of them \citep{datta2026erroneous}. Precision-first is
therefore a design priority of this pipeline and a direction of improvement,
not an achieved level of accuracy.

\subsection{The database used for modelling}
\label{sec:database}

The modelling results below use a consensus database that was assembled over
several staged runs during the campaign, as the corpus, the material vocabulary,
and the gate evolved. Its latest snapshot contains 448 consensus records over 22
inorganic materials from 39 contributing papers (\cref{tab:database}). It spans
perovskite ferroelectrics, niobates and tantalates, phosphate and iodate
crystals, a lead-titanate relaxor family, and a still-thin semiconductor set,
across electro-optic, refractive, dielectric, piezoelectric, and elastic
families. Three papers that recurred under different file names were detected
as byte-identical and de-duplicated.

\begin{table}[tb]
  \caption{Composition of the modelling database at its latest snapshot (448
  consensus records, 22 materials, 39 contributing papers). Counts cover all
  property families; every record carries a provenance tag and a confidence
  weight. Abbreviations: KTP, potassium titanyl phosphate; PMN--PT,
  Pb(Mg$_{1/3}$Nb$_{2/3}$)O$_3$--PbTiO$_3$; PZN--PT,
  Pb(Zn$_{1/3}$Nb$_{2/3}$)O$_3$--PbTiO$_3$; SBN, strontium barium niobate;
  PZT, lead zirconate titanate.}
  \label{tab:database}
  \centering
  \small
  \begin{tabular}{L{0.30\linewidth} r @{\hspace{2.5em}} L{0.30\linewidth} r}
    \toprule
    \DDHeadRow
    \DDHead{Material} & \DDHead{Records} & \DDHead{Material} & \DDHead{Records} \\
    \midrule
    BaTiO$_3$                       & 149 & Sr$_x$Ba$_{1-x}$Nb$_2$O$_6$ (SBN) & 9 \\
    BiFeO$_3$                       &  61 & AlN                               & 7 \\
    LiNbO$_3$                       &  47 & CdTe                              & 6 \\
    KNbO$_3$                        &  40 & Pb(Zr,Ti)O$_3$ (PZT)              & 6 \\
    PMN--PT                         &  30 & SrTiO$_3$                         & 3 \\
    LiIO$_3$                        &  20 & Ba$_x$Sr$_{1-x}$TiO$_3$           & 2 \\
    LiTaO$_3$                       &  19 & KTa$_{1-x}$Nb$_x$O$_3$            & 2 \\
    KTiOPO$_4$ (KTP)                &  17 & K$_3$Nb$_3$B$_2$O$_{12}$          & 1 \\
    PZN--PT                         &  13 & GaP                               & 1 \\
    BiScO$_3$--BiGaO$_3$--PbTiO$_3$ &  12 & InP                               & 1 \\
                                    &     & SiC                               & 1 \\
                                    &     & ZnTe                              & 1 \\
    \bottomrule
  \end{tabular}
\end{table}

\begin{Reproducibility}[Data lineage]
The September 2026 run of \cref{sec:run} is a single pass of the pipeline code
over the 49 PDFs and can be repeated with the same configuration. The
448-record database of \cref{tab:database} combines records from that run with
records from earlier staged runs, including runs with vocabulary extensions for
the relaxor solid solutions (PZN--PT, BiScO$_3$--BiGaO$_3$--PbTiO$_3$) and
ZnTe. Those earlier runs used earlier versions of the gate, which is why the
snapshots below are not directly comparable.
\end{Reproducibility}

\subsection{Coverage drives out-of-distribution generalization}
\label{sec:lomo}

As the database came to cover more distinct materials, the surrogate's ability
to predict held-out materials grew with it. \Cref{tab:trajectory} records four
snapshots taken during the campaign and \cref{fig:trajectory} plots them. The
number of materials whose LOMO error beats their baseline (\cref{eq:lomo}) rose
$1\to6\to8\to9$ as the number of materials rose $10\to15\to19\to22$, and the
share of the majority material (BaTiO$_3$) fell from 56\% to a third. The
fraction of tested materials that generalize rose from 25\% to 80\% and then
settled at 69\%, because each new material is often a single-source, data-poor
one that also joins the test pool.

The snapshots are not a controlled experiment. In a controlled experiment, one
factor (here, the number of materials) would be varied while everything else
was held fixed. Between our snapshots several things changed at once: the
papers in the corpus, the material vocabulary, and the version of the gate. The
rise in \cref{tab:trajectory} is therefore consistent with coverage driving
generalization, but it does not prove it, because a better gate or different
papers could contribute to the same rise.

\begin{table}[tb]
  \caption{The coverage--generalization trajectory over four database snapshots.
  The LOMO denominator grows because each snapshot admits new, often data-poor
  materials into the test pool. CV MAE (cross-validated mean absolute error) and
  baseline are in $\log_{10}$ units of the electro-optic coefficient. The
  snapshots differ in corpus, vocabulary, and gate version.}
  \label{tab:trajectory}
  \centering
  \small
  \begin{tabular}{l r r r c r r}
    \toprule
    \DDHeadRow
    \DDHead{Snapshot} & \DDHead{Records} & \DDHead{Materials} & \DDHead{BaTiO$_3$ (\%)} &
    \DDHead{LOMO beats} & \DDHead{CV MAE} & \DDHead{Baseline} \\
    \midrule
    Early   & 367 & 10 & 56 & 1/4   & 0.600 & 0.603 \\
    Merge   & 517 & 15 & 41 & 6/8   & 0.450 & 0.640 \\
    Rebuilt & 400 & 19 & 34 & 8/10  & 0.337 & 0.603 \\
    Current & 448 & 22 & 33 & 9/13  & 0.348 & 0.608 \\
    \bottomrule
  \end{tabular}
\end{table}

\begin{figure}[tb]
  \centering
  \begin{tikzpicture}
    \begin{axis}[ddlabj,
      xlabel={Materials in database}, ylabel={Held-out materials},
      xmin=8, xmax=24, ymin=0, ymax=14, ytick={0,2,...,14},
      legend pos=north west, legend cell align=left]
      \addplot+[thick] coordinates {(10,1)(15,6)(19,8)(22,9)};
      \addlegendentry{Generalize (LOMO beats baseline)}
      \addplot+[thick, dashed] coordinates {(10,4)(15,8)(19,10)(22,13)};
      \addlegendentry{Tested}
    \end{axis}
  \end{tikzpicture}
  \caption{Out-of-distribution generalization tracks the number of distinct
  materials across the four snapshots of \cref{tab:trajectory}. The count of
  held-out materials that beat their baseline rises with the number of
  materials; the test pool (dashed) grows in step, which is why the fraction
  does not rise monotonically.}
  \label{fig:trajectory}
\end{figure}

\begin{KeyFinding}
In these snapshots, generalization to unseen materials followed the number of
distinct materials rather than the number of records. The Merge snapshot had
the most records but fewer materials than the Current one; the early database
was dominated by BaTiO$_3$ and its model beat the baseline on one of four
held-out materials. A controlled test, which subsamples materials from a fixed
database, is needed to separate the effect of coverage from that of the changes
to the pipeline between snapshots.
\end{KeyFinding}

At the current snapshot the surrogate reaches a CV MAE of 0.35 $\log_{10}$
units against a predict-the-mean baseline of 0.61, and nine of thirteen
held-out materials beat their baseline (\cref{tab:lomo}). They span perovskites
(BaTiO$_3$, SrTiO$_3$), a ferrite (BiFeO$_3$), a niobate (SBN), a nitride
(AlN), a tantalate (LiTaO$_3$), and the lead-titanate relaxor family (PMN--PT,
PZN--PT, and the BiScO$_3$--BiGaO$_3$--PbTiO$_3$ solid solution).

\begin{table}[tb]
  \caption{Leave-one-material-out generalization at the current snapshot.
  $n$ is the number of electro-optic records of the held-out material entering
  the LOMO fit (\cref{tab:database} counts all families). \Term{Beat} means the
  held-out MAE is below the material's own predict-the-mean baseline
  (\cref{eq:lomo}). Errors are in $\log_{10}$ units.}
  \label{tab:lomo}
  \centering
  \small
  \begin{tabular}{l r r r l}
    \toprule
    \DDHeadRow
    \DDHead{Held-out material} & \DDHead{$n$} & \DDHead{LOMO MAE} & \DDHead{Baseline} & \DDHead{Verdict} \\
    \midrule
    LiTaO$_3$                       & 19 & 0.155 & 0.319 & Beat \\
    SrTiO$_3$                       &  3 & 0.160 & 1.157 & Beat \\
    PMN--PT                         &  4 & 0.239 & 0.867 & Beat \\
    BiScO$_3$--BiGaO$_3$--PbTiO$_3$ &  4 & 0.236 & 0.477 & Beat \\
    SBN                             &  5 & 0.292 & 0.325 & Beat \\
    PZN--PT                         &  8 & 0.520 & 0.791 & Beat \\
    BaTiO$_3$                       & 92 & 0.534 & 0.755 & Beat \\
    BiFeO$_3$                       & 48 & 0.810 & 1.188 & Beat \\
    AlN                             &  6 & 0.846 & 1.577 & Beat \\
    LiNbO$_3$                       & 18 & 0.341 & 0.293 & Miss \\
    KTP                             & 11 & 0.340 & 0.224 & Miss \\
    KNbO$_3$                        & 11 & 0.464 & 0.344 & Miss \\
    CdTe                            &  3 & 0.846 & 0.420 & Miss \\
    \bottomrule
  \end{tabular}
\end{table}

\subsection{What the gate caught}
\label{sec:caught}

Six interventions show the gate and the deterministic parsers acting on real
extraction output. Each is a value that a recall-first pipeline would have
entered wrong, mislabelled, or without indices.
\begin{itemize}[leftmargin=1.2em, itemsep=0.2em]
  \item \textbf{Symmetry rejection.} In the September run, a paper on
        GaInAsP/InP modulators \citep{krauser1986electro} yielded $r_{11}$,
        $r_{13}$, $r_{22}$, and $r_{33}$ for InP. InP is cubic zinc-blende,
        which permits only $r_{41}$ (\cref{eq:zb}); all four were rejected
        (\cref{lst:gate}), and the paper's only accepted record is a
        refractive index.
  \item \textbf{A number look-alike.} In an earlier staged run, a CdTe entry
        from a study of Zn$_x$Cd$_{1-x}$Te crystals \citep{jeon2007electro}
        read $r_{41}=\SI{110}{pm/V}$, far above any reported CdTe value.
        Tracing it to the source showed that ``110'' was the $\langle110\rangle$
        orientation along which the samples were cleaved. The record was
        removed; in the September run the same paper contributes the value it
        reports, $r_{41}=\SI{15.5}{pm/V}$.
  \item \textbf{Symbol recovery.} The $\gamma\to r$ rule recovered a
        five-component tensor from the hydrothermal KTiOPO$_4$ paper
        \citep{zhou2023hydrothermal} that a literal reader would have
        discarded as unrecognised labels (\cref{sec:gate}).
  \item \textbf{Identity separation.} A paper on
        0.88\,Pb(Zn$_{1/3}$Nb$_{2/3}$)O$_3$--0.12\,PbTiO$_3$ \citep{lu2000linear}
        was, before the vocabulary was extended, attributed to PbTiO$_3$
        through its end-member formula. Adding PZN--PT as a distinct identity
        kept its large coefficient ($r_{33}=\SI{134}{pm/V}$) from being
        assigned to PbTiO$_3$.
  \item \textbf{Unit consistency.} In the BaTiO$_3$ reference tensor set
        \citep{zgonik1994dielectric}, permittivities of about 4400 were read
        into elastic-constant slots. They were held instead of entering as
        elastic constants, because their source text carried no elastic unit.
  \item \textbf{Structured-tensor recovery.} A first-principles study of
        BaTiO$_3$ \citep{paoletta2021pockels} printed its Pockels tensors as
        bracketed $6\times3$ arrays of bare numbers, one per lattice constant.
        The deterministic parser assigned each entry its $r_{ij}$ label from
        its position.
\end{itemize}

\subsection{Limitations}
\label{sec:limits}

\begin{Limitation}
\begin{itemize}[leftmargin=1.2em, itemsep=0.15em, topsep=0.1em]
  \item \textbf{Accuracy.} Precision against the gold standard is about one
        half at full-record match (\cref{sec:accuracy}), and the LOMO test is
        only as trustworthy as the held-out labels it is scored against
        \citep{datta2026erroneous}.
  \item \textbf{Symmetry coverage.} Rejection of forbidden components is
        implemented for zinc-blende only. Extending it to every point group, with
        the phase and axis setting taken into account, is the cheapest
        improvement available.
  \item \textbf{Provenance mixing.} Measured and first-principles values enter
        one model, separated only by a feature; the electro-optic data contain
        more computed (90) than measured (67) values.
  \item \textbf{Weak baseline and small groups.} Predict-the-mean is a minimum
        bar, several held-out materials have three to five records, and LOMO was
        run once per snapshot. Per-component means, nearest neighbours in
        composition, repeated random seeds, and imputation inside each fold are
        needed for a firm estimate.
  \item \textbf{Semiconductors.} CdTe misses at $n=3$, and GaP, InP, SiC, and
        ZnTe contribute one record each, so the model has few zinc-blende
        neighbours. No pure PbTiO$_3$ single-crystal electro-optic measurement
        was found.
  \item \textbf{Resolution.} An MAE of about 0.35 $\log_{10}$ units is roughly a
        factor of two, adequate for ranking but not for quantitative design.
\end{itemize}
\end{Limitation}

\section{Discussion and outlook}
\label{sec:discussion}

\subsection{From a forward model to inverse design}
\label{sec:inverse}

The forward surrogate is not the end product; it is the operator that the
framework's generative stage inverts. \Cref{eq:designproblem} asks for the
material configuration and operating conditions that maximise the
electro-optic response. A model that generalizes across materials, and reports
where it does not, turns that question into a constrained search:

\begin{GoverningEquation}{Physics-constrained inverse design}[the generative target]
\begin{equation}\label{eq:design}
  \vect{x}^{\star} = \argmin_{\vect{x}\in\Set{X}}
  \Bigl[\,\Set{L}\bigl(f_\theta(\vect{x}),\, r_{\mathrm{target}}\bigr)
        + \beta_\sigma\,\sigma_\theta(\vect{x}) + \lambda\,\Phi_{\mathrm{phys}}(\vect{x})\Bigr],
  \qquad \vect{x}\in\Set{D}_\tau ,
\end{equation}
\end{GoverningEquation}

\noindent
where $\vect{x}=(\vect{c},\vect{k})$ joins the material configuration and the
operating conditions of \cref{eq:designproblem}, $\Set{X}$ is the space of
candidates, $r_{\mathrm{target}}$ the desired Pockels coefficient, $\Set{L}$ a
loss that is zero once the target is reached, $\sigma_\theta$ the ensemble
spread of \cref{sec:model}, which penalises extrapolation into data-poor regions
with weight $\beta_\sigma$, $\Phi_{\mathrm{phys}}$ a penalty with weight
$\lambda$ on charge imbalance, symmetry incompatibility, and synthesis
infeasibility, and $\Set{D}_\tau$ the domain over which the model has
demonstrated LOMO generalization. The database enters this expression twice: it
defines $f_\theta$, and its per-material LOMO verdicts define $\Set{D}_\tau$. A
candidate proposed outside that domain is flagged by $\sigma_\theta$ rather than
trusted.

\paragraph{What the inverse problem actually asks}
For quantum and integrated photonics the target is rarely ``the largest $r$''.
A cryogenic modulator or transducer needs a large $n^{3}r_{\mathrm{eff}}$ at a
few kelvin, low optical loss, and a dielectric constant that does not make the
electrodes impractically capacitive. The natural objective is therefore a
figure of merit evaluated at the operating point, for example
\begin{equation}\label{eq:fom}
  \mathrm{FOM}(\vect{c},\vect{k}) =
  \frac{n^{3}(\vect{c},\vect{k})\, r_{\mathrm{eff}}(\vect{c},\vect{k})}
       {\varepsilon_{r}(\vect{c},\vect{k})^{\,q}},
\end{equation}
where $\varepsilon_r$ is the relative permittivity and $q$ ($0\le q\le1$) sets
how strongly capacitance is penalised, which depends on the device. The
database already records refractive index, dielectric constants, and
electro-optic coefficients for the same materials, which is what
\cref{eq:fom} needs. Because the knobs $\vect{k}$ are inputs of $f_\theta$,
the same model can be asked a second kind of question: not which new
composition to make, but at which temperature, strain, or bias field an
existing crystal performs best. For barium titanate on a chosen substrate, for
example, the question becomes which strain state maximises
\cref{eq:fom} at \SI{4}{K}, the regime that strained-film experiments are now
probing \citep{suceava2026colossal, ulrich2025engineering}.

\paragraph{A closed loop}
\Cref{alg:loop} states the intended loop. The surrogate proposes candidates
inside $\Set{D}_\tau$; first-principles calculations of the Pockels tensor,
of the kind already in the corpus \citep{paoletta2021pockels, kim2023nature,
hamze2020design}, screen them; the survivors are synthesised and measured; and
every result, confirmed or not, returns to the database through the same gate,
so that the next round of the model learns from it. The ensemble spread
$\sigma_\theta$ tells Stage~0 which material or condition to look for in the
literature next \citep{lookman2019active}.

\begin{DDAlgorithm}[tb]{Closed-loop design over materials and operating conditions (planned)}{alg:loop}
\Require database $\Set{B}$; surrogate $f_\theta$ and ensemble spread
  $\sigma_\theta$; trusted domain $\Set{D}_\tau$; target $r_{\mathrm{target}}$
\Ensure a ranked list of confirmed configurations $(\vect{c},\vect{k})$
\Repeat
  \State propose $\{\vect{x}_j\}$ by solving \cref{eq:design} within $\Set{D}_\tau$
  \State screen each $\vect{x}_j$ with first-principles Pockels calculations
  \State synthesise and measure the survivors at their operating conditions $\vect{k}_j$
  \State pass every result through the gate (\cref{alg:gate}) into $\Set{B}$
  \State query Stage~0 for literature where $\sigma_\theta$ is largest
  \State refit $f_\theta$ and $\sigma_\theta$; recompute $\Set{D}_\tau$ by LOMO
\Until{a configuration meets $r_{\mathrm{target}}$ and is confirmed experimentally}
\end{DDAlgorithm}

\paragraph{What has to improve first}
Three gaps separate the present model from this loop. Its resolution, about a
factor of two (\cref{sec:limits}), is enough to rank candidates but not to
promise a value. Only about a fifth of the electro-optic records carry
complete temperature-and-strain context, so the condition dependence that
\cref{eq:fom} relies on is learned from few examples. And measured and computed
values are mixed in one model. Each gap is addressed by data rather than by a
different algorithm, which is why the next steps below concern the database.

\begin{PhysicsInsight}
The relaxor cluster is the concrete payoff. Because PMN--PT, PZN--PT, and the
BiScO$_3$--BiGaO$_3$--PbTiO$_3$ solid solution generalize together, the model
has learned a composition--property relation across that family rather than
three isolated look-ups. A mutually supporting cluster of this kind is where
\cref{eq:design} can first be inverted to propose a new relaxor composition at a
specified temperature and strain.
\end{PhysicsInsight}

\subsection{Next steps}
\label{sec:outlook}

\paragraph{Scale the corpus, with acquisition held to the same standard}
A first pass of Stage~0 assembled a catalogue of 1{,}032 records and 645
downloaded PDFs (560 unique). A screen with the pipeline's own material
detector and a search for coefficients reported in pm/V, which costs nothing
to run, found that 57 of the 557 papers not already in the corpus report
electro-optic coefficient values. Acquisition precision, not extraction cost,
is therefore the first constraint on scale, and the coefficient screen will be
applied as a gate before download.

\paragraph{Close the semiconductor gap}
A few measured zinc-blende and wurtzite sources (ZnTe, GaAs, GaN) would give GaP
and CdTe the chemical neighbours the composition features need.

\paragraph{Uncertainty-guided acquisition}
The modelling stage already emits what an active-learning loop consumes: a
per-material generalization verdict and a per-record confidence. Coupling the
ensemble spread $\sigma_\theta$ of \cref{eq:design} back into the relevance
ranker would let the system acquire the material, condition, or measurement
that most reduces its own uncertainty \citep{lookman2019active}.

\paragraph{Condition-resolved modelling}
As coverage of conditions grows, the surrogate can move from a composition
model toward the full $r(\vect{c}, T, \varepsilon, E_{\mathrm{bias}},
\lambda)$ response that \cref{eq:designproblem} targets. Cryogenic data are the
priority, because they are the regime of quantum transduction and the one
handbooks cover least.

\paragraph{Domain transfer}
Nothing in the five-stage architecture is specific to electro-optics. The
symmetry clause, the symbol-recovery rule, and the provenance weighting are
instances of one pattern, encoding a domain's physics as acceptance
constraints, that transfers to battery, catalysis, and quantum-materials
databases by substituting the vocabulary and the admissibility rules.

\begin{OpenQuestion}
Does uncertainty-guided acquisition, run as a closed loop against
\cref{eq:design}, find a new high-Pockels configuration that is then confirmed
by first-principles calculation and synthesis? That experiment is the reason
the framework exists, and it is the subject of ongoing work.
\end{OpenQuestion}

\begin{Reproducibility}
Every accepted record traces to its source publication and page, its extraction
route (LLM or deterministic parser), and its verification verdict, so a reader
can reconstruct why any value is in the database, or why a held value is not.
The per-stage accepted, review, and rejected tables of the September 2026 run,
with typed reasons, are available from the author on request.
\end{Reproducibility}

\begin{Declarations}
\AuthorContributions{%
  D.D.: conceptualisation, methodology, software, validation, formal analysis,
  investigation, data curation, writing}
\Funding{%
  This work is supported by the National Science Foundation (NSF) Designing
  Materials to Revolutionize and Engineer our Future (DMREF) Program (award
  number 2522898). Any opinions, findings, and conclusions expressed are those
  of the author and do not necessarily reflect the views of the National Science
  Foundation.}
\DataAvailability{%
  The record files of the September 2026 production run (accepted, review, and
  rejected, with typed reasons) and the gold-standard records are available
  from the author on reasonable request. A public repository with a DOI will be
  registered at journal submission.}
\CodeAvailability{The pipeline code is available from the author on request
  and will be released at \DDJCodeLink.}
\CompetingInterests{The author declares no competing interests.}
\AIDisclosure{%
  A large language model (GPT-4o, OpenAI) performed context-dependent extraction
  of candidate property values within the pipeline of \cref{sec:extract}; every
  extracted value passed the deterministic verification gate, and its accuracy
  is reported in \cref{sec:accuracy}. All quantitative
  claims, symmetry rules, and generalization results were checked by the author (Dibakar Datta),
  who is responsible for all content.}
\Acknowledgments{%
  The author thanks the DDLab members for discussions, and acknowledges the
  Advanced Cyberinfrastructure Coordination Ecosystem: Services \& Support
  (ACCESS) for computational resources (allocations DMR180013 and MAT250009).}


\end{Declarations}

\bibliographystyle{unsrtnat}
\bibliography{references}

\DDAppendices
\section{The end-to-end procedure}
\label{app:procedure}

The pipeline of \cref{fig:pipeline} is stated once as a procedure, so that the
manuscript and the code agree on order and on where a value is allowed to stop.

\begin{DDProcedure}{Precision-first knowledge generation and generalization test}{alg:procedure}
\PInput{A set of candidate publications; the material vocabulary $\Set{M}$; the
family label patterns; the symmetry and scope rules of \cref{lst:gate}.}
\PStep{Acquire} Query the bibliographic services, de-duplicate, score each
record with \cref{eq:relevance}, and download and validate papers above the
threshold.
\PStep{Ingest and de-duplicate} Parse each PDF into ordered, layout-aware text
and tables. Detect byte-identical duplicates and keep one copy.
\PStep{Detect} Identify the subject material with the confidence guard and gated
abbreviations of \cref{alg:detect}; record paper-level form, substrate,
temperature, and frequency. Hold papers whose material is uncertain and route
out-of-scope documents to a held list.
\PStep{Extract} Recover candidate records with the deterministic table parsers
and the LLM, each bound to a material, tensor label, unit, result type, and
conditions; prefer the deterministic row when both report the same quantity.
\PStep{Verify} Apply the three-tier gate of \cref{alg:gate}: recover symbols
($\gamma\to r$, boundary superscripts), reject symmetry-forbidden and
out-of-scope values, fold symmetry-equivalent labels, tag provenance, and
require a recognised label and a unit consistent with the family. Emit
\Term{accepted}, or hold or reject with a typed reason.
\PStep{Reconcile} Group accepted records and apply \cref{alg:consensus};
assign confidence weights; keep single-source records at reduced weight and move
outliers to review. Never delete.
\PStep{Model and test} Train the noise-robust surrogate with the weights of
\cref{eq:weight}; run the out-of-fold label audit and the leave-one-material-out
test (\cref{eq:lomo}).
\POutput{A canonical, weighted, traceable property database; a per-material
generalization verdict; and a flagged-record list for review.}
\PNote{Every rejected and held value is logged with its reason. A database that
records only accepted values cannot be audited, and cannot tell the next
acquisition step what it is missing.}
\end{DDProcedure}

\section{The corpus}
\label{app:corpus}

\Cref{tab:corpus} lists the 48 distinct works in the 49-file corpus of the
production run (one work appears both as a preprint and as its published
version). Papers marked $\dagger$ belong to the gold standard of
\cref{sec:gold}; the last three, on organic and hybrid
materials, are outside the scope of the database.

{\small
\begin{longtable}{L{0.36\linewidth} L{0.42\linewidth} l}
  \caption{The papers mined in this study.}\label{tab:corpus}\\
  \toprule
  \DDHeadRow
  \DDHead{Material} & \DDHead{Topic} & \DDHead{Reference} \\
  \midrule
  \endfirsthead
  \toprule
  \DDHeadRow
  \DDHead{Material} & \DDHead{Topic} & \DDHead{Reference} \\
  \midrule
  \endhead
  \bottomrule
  \endfoot
  BaTiO$_3$$^{\dagger}$ & reference tensor set, bulk crystal & \citep{zgonik1994dielectric} \\
  KNbO$_3$$^{\dagger}$ & reference tensor set, bulk crystal & \citep{zgonik1993materials} \\
  LiNbO$_3$$^{\dagger}$ & reference tensor set, bulk crystal & \citep{jazbinsek2002material} \\
  LiNbO$_3$ (Er, In:Er doped) & $r_{13}$, $r_{33}$ of doped crystals & \citep{du2017electro} \\
  LiTaO$_3$$^{\dagger}$ & near-infrared coefficients & \citep{casson2004electro} \\
  LiTaO$_3$ (near-stoichiometric) & deep-ultraviolet electro-optics & \citep{juvalta2007near} \\
  BaTiO$_3$$^{\dagger}$ & first principles, tetragonal phase & \citep{kim2023nature} \\
  BaTiO$_3$$^{\dagger}$ & first principles, Pockels matrices vs.\ lattice constant & \citep{paoletta2021pockels} \\
  BaTiO$_3$$^{\dagger}$ & strained thin film, cryogenic & \citep{suceava2026colossal} \\
  BaTiO$_3$, LiNbO$_3$$^{\dagger}$ & thin film, MHz to sub-THz response & \citep{chelladurai2025barium} \\
  BaTiO$_3$ & thin film on silicon & \citep{abel2019large} \\
  BaTiO$_3$ & single-crystalline thin films & \citep{wen2024enhanced} \\
  BaTiO$_3$ & thin-film tensor characterisation & \citep{cao2023characterization} \\
  BaTiO$_3$ & films on Si, thermal-rate control & \citep{reynaud2025enhancement} \\
  BaTiO$_3$ (doped) & nanocrystals & \citep{wang2022engineering} \\
  Lead-free perovskites & thin films, Pockels and Kerr & \citep{ion2023lead} \\
  Ba$_x$Sr$_{1-x}$TiO$_3$ & electro-optic filter at 2\,\textmu m & \citep{vakilitabatabaei2026electro} \\
  SrTiO$_3$$^{\dagger}$ & thin film, cryogenic (preprint and published version) & \citep{ulrich2025engineering} \\
  Perovskite oxides & design rules, first-principles screening & \citep{hamze2020design} \\
  BiFeO$_3$$^{\dagger}$ & thin-film optical modulators & \citep{zhu2016ferroelectric} \\
  BiFeO$_3$ & thin films, linear electro-optic effect & \citep{sando2014linear} \\
  PIN--PMN--PT$^{\dagger}$ & single-domain crystal & \citep{liu2024large} \\
  Sm-doped PIN--PMN--PT & domain-engineered crystal & \citep{wen2024giant} \\
  Pb(Zr,Ti)O$_3$ & plasmonic modulator, $r_{31}$ & \citep{yezekyan2024plasmonic} \\
  Pb(Zr,Ti)O$_3$ & thin-film Pockels laser & \citep{zhang2026ultrafast} \\
  Pb(Zr,Ti)O$_3$ (Ni-doped) & thin films & \citep{zhu2018strong} \\
  KTa$_{1-x}$Nb$_x$O$_3$$^{\dagger}$ & crystal growth and electro-optics (review) & \citep{imai2007crystal} \\
  KTa$_{1-x}$Nb$_x$O$_3$ & microresonator tuning & \citep{szabados2020electro} \\
  Sr$_x$Ba$_{1-x}$Nb$_2$O$_6$ & $r_{33}$ vs.\ temperature & \citep{goulkov2003temperature} \\
  Sr$_x$Ba$_{1-x}$Nb$_2$O$_6$ & thin films (book chapter) & \citep{cuniotponsard2011strontium} \\
  KTiOPO$_4$ & hydrothermal crystal, full tensor & \citep{zhou2023hydrothermal} \\
  Na:KTiOPO$_4$, RbTiOAsO$_4$ & low-frequency coefficients & \citep{xie2005low} \\
  K$_3$Nb$_3$B$_2$O$_{12}$$^{\dagger}$ & new electro-optic crystal & \citep{lv2025new} \\
  $\beta$-BaB$_2$O$_4$ & Q-switch & \citep{goodno1995investigation} \\
  LiIO$_3$ & first-principles optical properties & \citep{vantroeye2014first} \\
  KH$_2$PO$_4$ (Mg-doped) & first-principles optical properties & \citep{zhu2024first} \\
  AlN & bulk nonlinear and electro-optic coefficients & \citep{majkic2017optical} \\
  ScAlN & Pockels coefficient of a ferroelectric nitride & \citep{yang2024unveiling} \\
  SiC (3C)$^{\dagger}$ & integrated modulator & \citep{powell2022integrated} \\
  GaAs/AlGaAs & ultralow-$V_\pi$ modulator & \citep{shin2013ultralow} \\
  GaAs & push-pull modulator, $r_{41}$ & \citep{li2025engineering} \\
  InP, GaInAsP & electro-optic modulators & \citep{krauser1986electro} \\
  GaP$^{\dagger}$ & dispersion of $r_{41}$ & \citep{berozashvili1989dispersion} \\
  Zn$_x$Cd$_{1-x}$Te$^{\dagger}$ & single crystals at 1550\,nm & \citep{jeon2007electro} \\
  (Cd,Mg)Te & single crystals & \citep{serafini2017photoconductive} \\
  Metal-free perovskites & electro-optic modulation (out of scope) & \citep{gao2021electro} \\
  Hybrid halide perovskites & electro-optic modulation (out of scope) & \citep{gao2019electro} \\
  Conjugated molecules & organic electro-optics (out of scope) & \citep{gao2021self} \\
\end{longtable}}

\end{document}

%% file: Preamble/DDLabIdentity.tex
\labname{Dibakar Datta Laboratory}
\labshortname{DDLab}
\institutionshort{NJIT}

\labwebsite[https://dibakardatta.net/]{www.dibakardatta.net}

\labcodeurl{https://dibakardatta.net/code/}

\citationauthors{D. Datta \emph{et al.}}

%% file: Figures/fig_eo_physics.tex
\begin{figure}[tbp]
  \centering
  \begin{tikzpicture}[x=1mm, y=1mm, font=\footnotesize,
      every node/.append style={execute at begin node={\hyphenpenalty=10000\relax}},
      mw/.style={DDNavy, line width=0.9pt, decorate,
                 decoration={snake, amplitude=1.6mm, segment length=7mm}},
      opt/.style={DDPurple, line width=0.9pt, decorate,
                  decoration={snake, amplitude=1.1mm, segment length=2.6mm}},
      lab/.style={font=\scriptsize, text=DDMeta, align=center},
      head/.style={font=\footnotesize\bfseries, align=center}]

    \node[anchor=west, font=\small\bfseries, text=DDNavy] at (0,64) {(a)};
    \node[anchor=west, font=\small, text=DDInk] at (6,64)
      {The electro-optic link of a networked quantum processor};

    \draw[DDNavy!70, dashed, rounded corners=2mm, fill=DDNavy!3]
      (0,8) rectangle (97,58);
    \node[anchor=north west, font=\scriptsize\bfseries, text=DDNavy]
      at (1.5,57) {Cryostat: millikelvin to a few kelvin};

    \draw[DDNavy, fill=white, rounded corners=1mm] (5,20) rectangle (31,44);
    \foreach \x in {10,18,26} {
      \draw[DDNavy, line width=0.7pt] (\x-2.2,32-2.2) -- (\x+2.2,32+2.2);
      \draw[DDNavy, line width=0.7pt] (\x-2.2,32+2.2) -- (\x+2.2,32-2.2);
      \draw[DDNavy, line width=0.7pt] (\x,26) -- (\x,29.8);
      \draw[DDNavy, line width=0.7pt] (\x,34.2) -- (\x,38);
    }
    \node[head, text=DDNavy] at (18,48.5) {Qubits};
    \node[lab] at (18,15) {superconducting\\circuits, $\sim$10\,mK};

    \draw[mw] (32,32) -- (51,32);
    \draw[-{Stealth[length=2mm]}, DDNavy] (51,32) -- (54,32);
    \node[lab, text=DDNavy] at (43,40) {microwave\\photon, $\sim$5\,GHz};

    \fill[DDTeal!10] (55,24) rectangle (89,40);
    \draw[DDTeal, line width=0.6pt] (55,24) rectangle (89,40);
    \fill[DDGold!65] (57,40) rectangle (87,42.2);
    \fill[DDGold!65] (57,21.8) rectangle (87,24);
    \foreach \x in {61,67,...,85} {
      \draw[-{Stealth[length=1.4mm]}, DDOrange!85, line width=0.5pt] (\x,39) -- (\x,34);
    }
    \node[font=\scriptsize, text=DDTeal!80!black] at (72,29.5)
      {electro-optic crystal, $r_{ij}$};
    \node[font=\scriptsize, text=DDOrange!85!black, anchor=west] at (88.5,37.5) {$E$};
    \node[head, text=DDTeal!80!black] at (72,48.5) {Transducer};
    \node[lab] at (68,15) {$\Delta(1/n^{2})_i = r_{ij}E_j$ mixes the\\microwave
      field into the light};

    \draw[-{Stealth[length=2mm]}, DDRed!80, line width=0.8pt] (85,2) -- (85,21.5);
    \node[lab, text=DDRed!80!black, anchor=west] at (86,4) {optical pump};

    \draw[opt] (90,32) -- (122,32);
    \draw[-{Stealth[length=2mm]}, DDPurple] (122,32) -- (125,32);
    \node[lab, text=DDPurple] at (110,40) {optical photon,\\$\sim$194\,THz (1550\,nm)};
    \node[lab] at (110,22) {low-loss fibre,\\room temperature};

    \draw[DDPurple, fill=DDPurple!5, rounded corners=1mm] (126,20) rectangle (163,44);
    \node[head, text=DDPurple] at (144.5,48.5) {Optical network};
    \node[lab, text=DDInk] at (144.5,32) {other quantum\\processors,\\memories, detectors};

    \begin{scope}[yshift=-66mm]
    \node[anchor=west, font=\small\bfseries, text=DDNavy] at (0,55) {(b)};
    \node[anchor=west, font=\small, text=DDInk] at (6,55)
      {The Pockels effect: a field changes the index and shifts the phase};

    \draw[DDRed!80, line width=0.9pt, domain=0:34, samples=120, smooth]
      plot (\x, {34 + 5*sin(\x/6.5*360)});
    \node[lab, text=DDRed!80!black] at (17,45) {input light};
    \node[lab] at (17,24) {$E_0\cos(\omega t)$};
    \draw[-{Stealth[length=2mm]}, DDRed!80, line width=0.8pt] (35,34) -- (44,34);

    \fill[DDTeal!10] (45,26) rectangle (105,42);
    \draw[DDTeal, line width=0.6pt] (45,26) rectangle (105,42);
    \fill[DDGold!65] (45,42) rectangle (105,45);
    \fill[DDGold!65] (45,23) rectangle (105,26);
    \node[font=\scriptsize, text=DDGold!40!black] at (75,47.5) {electrode, $+V$};
    \node[font=\scriptsize, text=DDGold!40!black] at (75,20.5) {electrode, ground};
    \foreach \x in {50,58,...,98} {
      \draw[-{Stealth[length=1.4mm]}, DDOrange!85, line width=0.5pt] (\x,41) -- (\x,36);
    }
    \draw[DDRed!70, line width=0.6pt, dashed] (45,31) -- (105,31);
    \node[font=\scriptsize, text=DDTeal!80!black] at (75,28.3)
      {$n \;\rightarrow\; n + \Delta n$};
    \draw[{Stealth[length=1.6mm]}-{Stealth[length=1.6mm]}, DDMeta] (45,16) -- (105,16);
    \node[lab, fill=white, inner sep=1pt] at (75,16) {length $L$};
    \draw[{Stealth[length=1.4mm]}-{Stealth[length=1.4mm]}, DDMeta] (108,26) -- (108,42);
    \node[lab, anchor=west] at (109,34) {$d$};

    \draw[-{Stealth[length=2mm]}, DDRed!80, line width=0.8pt] (113,34) -- (122,34);
    \draw[DDRed!45, line width=0.7pt, dashed, domain=124:163, samples=120, smooth]
      plot (\x, {34 + 5*sin((\x-124)/6.5*360)});
    \draw[DDPurple, line width=0.9pt, domain=124:163, samples=120, smooth]
      plot (\x, {34 + 5*sin((\x-124)/6.5*360 - 70)});
    \node[lab, text=DDPurple] at (143.5,45) {output light};
    \node[lab] at (143.5,24) {$E_0\cos(\omega t + \Delta\varphi)$\\[-0.3mm]{\color{DDRed!60}dashed: without field}};

    \node[draw=DDRule, fill=DDLight, rounded corners=1mm, inner sep=2.5mm,
          font=\footnotesize, align=center] at (81.5,4)
      {$\Delta n = -\tfrac12\, n^{3} r_{\mathrm{eff}} E$, \quad
       $\Delta\varphi = \dfrac{2\pi}{\lambda}\,\Delta n\, L$, \quad
       $E = V/d$ \qquad$\Longrightarrow$\qquad
       larger $n^{3}r_{\mathrm{eff}}$: lower $V_\pi$, less energy per bit, shorter devices};
    \end{scope}
  \end{tikzpicture}
  \caption{Why electro-optic materials matter.
  (a)~If qubits are the brain of a quantum computer, electro-optic materials
  are its nervous system. Superconducting qubits work at millikelvin
  temperatures and speak in microwave photons, whereas low-loss transmission
  over distance uses optical photons. A cavity electro-optic transducer couples
  the two through the Pockels effect of a crystal held at cryogenic temperature
  \citep{lauk2020perspectives, han2021microwave, holzgrafe2020cavity,
  sahu2022quantum}; its efficiency and added noise depend on the Pockels
  coefficient the crystal keeps at a few kelvin.
  (b)~In a phase modulator the applied field $E=V/d$ changes the refractive
  index by $\Delta n$ (\cref{eq:pockels}) and shifts the phase of the light by
  $\Delta\varphi$. A larger $n^{3}r_{\mathrm{eff}}$ lowers the half-wave voltage
  $V_\pi$ (\cref{eq:vpi}) and the switching energy, which scales as $CV^{2}$
  for electrode capacitance $C$.}
  \label{fig:eo-physics}
\end{figure}

%% file: Figures/fig_design_space.tex
\begin{figure}[tbp]
  \centering
  \begin{tikzpicture}[x=1mm, y=1mm, font=\footnotesize,
      every node/.append style={execute at begin node={\hyphenpenalty=10000\relax}},
      lab/.style={font=\scriptsize, text=DDMeta, align=center},
      head/.style={font=\footnotesize\bfseries, align=center},
      panel/.style={draw=#1, fill=#1!4, rounded corners=1.5mm, line width=0.6pt},
      atA/.style={circle, fill=DDNavy!70, inner sep=0pt, minimum size=2.6mm},
      atB/.style={circle, fill=DDRed!75, inner sep=0pt, minimum size=2.4mm},
      atO/.style={circle, draw=DDGray, fill=white, inner sep=0pt, minimum size=1.6mm}]

    \draw[panel=DDNavy] (0,0) rectangle (66,62);
    \node[head, text=DDNavy] at (33,57.5) {Material configuration $\vect{c}$};

    \begin{scope}[shift={(17,36)}]
      \fill[DDRed!8] (0,8) -- (8,0) -- (0,-8) -- (-8,0) -- cycle;
      \draw[DDGray, line width=0.4pt] (0,8) -- (8,0) -- (0,-8) -- (-8,0) -- cycle;
      \draw[DDGray!70, line width=0.3pt] (-8,0) -- (8,0) (0,-8) -- (0,8);
      \draw[DDNavy!50, line width=0.4pt] (-11,-11) rectangle (11,11);
      \foreach \p in {(-11,-11),(11,-11),(11,11),(-11,11)} \node[atA] at \p {};
      \node[atB] at (0,0) {};
      \foreach \p in {(0,8),(8,0),(0,-8),(-8,0)} \node[atO] at \p {};
      \node[font=\scriptsize, text=DDNavy] at (-11,-14.5) {A};
      \node[font=\scriptsize, text=DDRed!80!black] at (3.2,2.6) {B};
      \node[font=\scriptsize, text=DDMeta] at (6.2,-3.6) {O};
    \end{scope}
    \node[lab] at (17,19.5) {perovskite ABO$_3$};

    \node[anchor=north west, align=left, font=\scriptsize, text=DDInk,
          text width=34mm] at (31,51)
      {\textbf{Simple:} ABO$_3$\\[1.2mm]
       \textbf{Solid solutions:}\\(A,A$'$)(B,B$'$)O$_3$\\[1.2mm]
       \textbf{Superlattices:}\\(ABO$_3$)$_m$/(A$'$B$'$O$_3$)$_n$\\[1.2mm]
       \textbf{Layered:} Ruddlesden--Popper, Dion--Jacobson, Aurivillius};
    \node[anchor=south, align=center, font=\scriptsize, text=DDInk,
          text width=62mm] at (33,2)
      {also niobates, tantalates, phosphates, nitrides, and zinc-blende
       semiconductors; phase, orientation, domain state, film or bulk};

    \node[font=\LARGE, text=DDMeta] at (70.5,31) {$\times$};

    \draw[panel=DDOrange] (75,0) rectangle (125,62);
    \node[head, text=DDOrange!85!black] at (100,57.5) {Operating knobs $\vect{k}$};

    \begin{scope}[shift={(83,44)}]
      \draw[DDOrange!85, line width=0.6pt, rounded corners=0.8mm] (-1.2,-3) rectangle (1.2,6);
      \fill[DDNavy!70] (0,-4.2) circle (2.1mm);
      \fill[DDNavy!70] (-0.6,-3) rectangle (0.6,0.5);
    \end{scope}
    \node[anchor=west, font=\scriptsize, text=DDInk, align=left] at (88,45)
      {temperature $T$:\\millikelvin to room temperature};

    \begin{scope}[shift={(83,29)}]
      \draw[DDTeal, fill=DDTeal!10] (-3,-3) rectangle (3,3);
      \draw[-{Stealth[length=1.2mm]}, DDTeal] (3.6,0) -- (6.2,0);
      \draw[-{Stealth[length=1.2mm]}, DDTeal] (-3.6,0) -- (-6.2,0);
      \draw[-{Stealth[length=1.2mm]}, DDTeal] (0,3.6) -- (0,6.2);
      \draw[-{Stealth[length=1.2mm]}, DDTeal] (0,-3.6) -- (0,-6.2);
    \end{scope}
    \node[anchor=west, font=\scriptsize, text=DDInk, align=left] at (90,29)
      {strain $\varepsilon$: biaxial epitaxial\\(substrate) or uniaxial};

    \begin{scope}[shift={(83,14)}]
      \fill[DDGold!65] (-4,3) rectangle (4,4.2);
      \fill[DDGold!65] (-4,-4.2) rectangle (4,-3);
      \foreach \x in {-2.5,0,2.5}
        \draw[-{Stealth[length=1.1mm]}, DDOrange!85] (\x,2.6) -- (\x,-2.6);
    \end{scope}
    \node[anchor=west, font=\scriptsize, text=DDInk, align=left] at (90,14)
      {bias field $E_{\mathrm{bias}}$ (can induce\\a linear response)};
    \node[anchor=south, font=\scriptsize, text=DDInk] at (100,2)
      {wavelength $\lambda$, modulation frequency $f$};

    \draw[-{Stealth[length=2.4mm]}, DDNavy, line width=1pt] (126.5,31) -- (133,31);

    \draw[panel=DDTeal] (134,0) rectangle (165,62);
    \node[head, text=DDTeal!80!black] at (149.5,57.5) {Response};
    \node[font=\normalsize, text=DDTeal!80!black] at (149.5,46)
      {$n^{3}r_{\mathrm{eff}}(\vect{c},\vect{k})$};
    \node[lab, text width=28mm] at (149.5,35)
      {target: maximum at the operating point, e.g.\ 4\,K and a given strain};
    \draw[DDRule] (137,25) -- (162,25);
    \node[lab, text width=29mm, text=DDRed!80!black] at (149.5,13.5)
      {The Materials Project and OQMD hold no $r_{ij}$ and no measurement conditions; the data exist only in papers.};
  \end{tikzpicture}
  \caption{The electro-optic design problem of \cref{eq:designproblem}. The
  response depends jointly on the material configuration, which for perovskite
  oxides alone spans solid solutions, superlattices, and layered phases, and on
  external knobs that can each change the Pockels coefficient substantially:
  temperature \citep{eltes2020integrated, ulrich2025engineering}, strain
  \citep{paoletta2021pockels, suceava2026colossal}, and bias field
  \citep{imai2007crystal}. The joint space is far too large to search by
  experiment alone, and a model that could search it needs condition-resolved
  Pockels data that no general materials database contains. Building those data
  is the subject of this paper.}
  \label{fig:design-space}
\end{figure}

%% file: references.bib
@article{jain2013materials,
  author  = {Jain, Anubhav and Ong, Shyue Ping and Hautier, Geoffroy and Chen, Wei and Richards, William Davidson and Dacek, Stephen and Cholia, Shreyas and Gunter, Dan and Skinner, David and Ceder, Gerbrand and Persson, Kristin A.},
  title   = {Commentary: The {Materials Project}: A materials genome approach to accelerating materials innovation},
  journal = {APL Materials},
  volume  = {1},
  number  = {1},
  pages   = {011002},
  year    = {2013},
  doi     = {10.1063/1.4812323}
}

@article{saal2013oqmd,
  author  = {Saal, James E. and Kirklin, Scott and Aykol, Muratahan and Meredig, Bryce and Wolverton, Christopher},
  title   = {Materials design and discovery with high-throughput density functional theory: The {Open Quantum Materials Database} ({OQMD})},
  journal = {JOM},
  volume  = {65},
  number  = {11},
  pages   = {1501--1509},
  year    = {2013},
  doi     = {10.1007/s11837-013-0755-4}
}

@article{curtarolo2012aflow,
  author  = {Curtarolo, Stefano and Setyawan, Wahyu and Hart, Gus L. W. and Jahnatek, Michal and Chepulskii, Roman V. and Taylor, Richard H. and Wang, Shidong and Xue, Junkai and Yang, Kesong and Levy, Ohad and others},
  title   = {{AFLOW}: An automatic framework for high-throughput materials discovery},
  journal = {Computational Materials Science},
  volume  = {58},
  pages   = {218--226},
  year    = {2012},
  doi     = {10.1016/j.commatsci.2012.02.005}
}

@article{belsky2002icsd,
  author  = {Belsky, Alec and Hellenbrandt, Mariette and Karen, Vicky Lynn and Luksch, Peter},
  title   = {New developments in the {Inorganic Crystal Structure Database} ({ICSD}): accessibility in support of materials research and design},
  journal = {Acta Crystallographica Section B},
  volume  = {58},
  number  = {3},
  pages   = {364--369},
  year    = {2002},
  doi     = {10.1107/S0108768102006948}
}

@article{tshitoyan2019unsupervised,
  author  = {Tshitoyan, Vahe and Dagdelen, John and Weston, Leigh and Dunn, Alexander and Rong, Ziqin and Kononova, Olga and Persson, Kristin A. and Ceder, Gerbrand and Jain, Anubhav},
  title   = {Unsupervised word embeddings capture latent knowledge from materials science literature},
  journal = {Nature},
  volume  = {571},
  number  = {7763},
  pages   = {95--98},
  year    = {2019},
  doi     = {10.1038/s41586-019-1335-8}
}

@article{olivetti2020data,
  author  = {Olivetti, Elsa A. and Cole, Jacqueline M. and Kim, Edward and Kononova, Olga and Ceder, Gerbrand and Han, Thomas Yong-Jin and Hiszpanski, Anna M.},
  title   = {Data-driven materials research enabled by natural language processing and information extraction},
  journal = {Applied Physics Reviews},
  volume  = {7},
  number  = {4},
  pages   = {041317},
  year    = {2020},
  doi     = {10.1063/5.0021106}
}

@article{kononova2021opportunities,
  author  = {Kononova, Olga and He, Tanjin and Huo, Haoyan and Trewartha, Amalie and Olivetti, Elsa A. and Ceder, Gerbrand},
  title   = {Opportunities and challenges of text mining in materials research},
  journal = {iScience},
  volume  = {24},
  number  = {3},
  pages   = {102155},
  year    = {2021},
  doi     = {10.1016/j.isci.2021.102155}
}

@article{swain2016chemdataextractor,
  author  = {Swain, Matthew C. and Cole, Jacqueline M.},
  title   = {{ChemDataExtractor}: A toolkit for automated extraction of chemical information from the scientific literature},
  journal = {Journal of Chemical Information and Modeling},
  volume  = {56},
  number  = {10},
  pages   = {1894--1904},
  year    = {2016},
  doi     = {10.1021/acs.jcim.6b00207}
}

@article{court2018auto,
  author  = {Court, Callum J. and Cole, Jacqueline M.},
  title   = {Auto-generated materials database of {Curie} and {N\'eel} temperatures via semi-supervised relationship extraction},
  journal = {Scientific Data},
  volume  = {5},
  pages   = {180111},
  year    = {2018},
  doi     = {10.1038/sdata.2018.111}
}

@article{huang2020database,
  author  = {Huang, Shu and Cole, Jacqueline M.},
  title   = {A database of battery materials auto-generated using {ChemDataExtractor}},
  journal = {Scientific Data},
  volume  = {7},
  pages   = {260},
  year    = {2020},
  doi     = {10.1038/s41597-020-00602-2}
}

@article{kim2017materials,
  author  = {Kim, Edward and Huang, Kevin and Saunders, Adam and McCallum, Andrew and Ceder, Gerbrand and Olivetti, Elsa},
  title   = {Materials synthesis insights from scientific literature via text extraction and machine learning},
  journal = {Chemistry of Materials},
  volume  = {29},
  number  = {21},
  pages   = {9436--9444},
  year    = {2017},
  doi     = {10.1021/acs.chemmater.7b03500}
}

@article{polak2024extracting,
  author  = {Polak, Maciej P. and Morgan, Dane},
  title   = {Extracting accurate materials data from research papers with conversational language models and prompt engineering},
  journal = {Nature Communications},
  volume  = {15},
  pages   = {1569},
  year    = {2024},
  doi     = {10.1038/s41467-024-45914-8}
}

@article{dagdelen2024structured,
  author  = {Dagdelen, John and Dunn, Alexander and Lee, Sanghoon and Walker, Nicholas and Rosen, Andrew S. and Ceder, Gerbrand and Persson, Kristin A. and Jain, Anubhav},
  title   = {Structured information extraction from scientific text with large language models},
  journal = {Nature Communications},
  volume  = {15},
  pages   = {1418},
  year    = {2024},
  doi     = {10.1038/s41467-024-45563-x}
}

@article{ji2023hallucination,
  author  = {Ji, Ziwei and Lee, Nayeon and Frieske, Rita and Yu, Tiezheng and Su, Dan and Xu, Yan and Ishii, Etsuko and Bang, Ye Jin and Madotto, Andrea and Fung, Pascale},
  title   = {Survey of hallucination in natural language generation},
  journal = {ACM Computing Surveys},
  volume  = {55},
  number  = {12},
  pages   = {248},
  year    = {2023},
  doi     = {10.1145/3571730}
}

@book{nye1985physical,
  author    = {Nye, J. F.},
  title     = {Physical Properties of Crystals: Their Representation by Tensors and Matrices},
  publisher = {Oxford University Press},
  address   = {Oxford},
  year      = {1985}
}

@book{iglewicz1993outliers,
  author    = {Iglewicz, Boris and Hoaglin, David C.},
  title     = {How to Detect and Handle Outliers},
  series    = {The ASQC Basic References in Quality Control: Statistical Techniques},
  volume    = {16},
  publisher = {ASQC Quality Press},
  address   = {Milwaukee, WI},
  year      = {1993}
}

@article{leys2013detecting,
  author  = {Leys, Christophe and Ley, Christophe and Klein, Olivier and Bernard, Philippe and Licata, Laurent},
  title   = {Detecting outliers: Do not use standard deviation around the mean, use absolute deviation around the median},
  journal = {Journal of Experimental Social Psychology},
  volume  = {49},
  number  = {4},
  pages   = {764--766},
  year    = {2013},
  doi     = {10.1016/j.jesp.2013.03.013}
}

@article{friedman2001greedy,
  author  = {Friedman, Jerome H.},
  title   = {Greedy function approximation: A gradient boosting machine},
  journal = {The Annals of Statistics},
  volume  = {29},
  number  = {5},
  pages   = {1189--1232},
  year    = {2001},
  doi     = {10.1214/aos/1013203451}
}

@article{huber1964robust,
  author  = {Huber, Peter J.},
  title   = {Robust estimation of a location parameter},
  journal = {The Annals of Mathematical Statistics},
  volume  = {35},
  number  = {1},
  pages   = {73--101},
  year    = {1964},
  doi     = {10.1214/aoms/1177703732}
}

@article{northcutt2021confident,
  author  = {Northcutt, Curtis G. and Jiang, Lu and Chuang, Isaac L.},
  title   = {Confident learning: Estimating uncertainty in dataset labels},
  journal = {Journal of Artificial Intelligence Research},
  volume  = {70},
  pages   = {1373--1411},
  year    = {2021},
  doi     = {10.1613/jair.1.12125}
}

@article{lookman2019active,
  author  = {Lookman, Turab and Balachandran, Prasanna V. and Xue, Dezhen and Yuan, Ruihao},
  title   = {Active learning in materials science with emphasis on adaptive sampling using uncertainties for targeted design},
  journal = {npj Computational Materials},
  volume  = {5},
  pages   = {21},
  year    = {2019},
  doi     = {10.1038/s41524-019-0153-8}
}

@misc{datta2026framework,
  author       = {Datta, Dibakar},
  title        = {An autonomous scientific knowledge generation framework for {AI}-driven scientific discovery},
  howpublished = {Preprint, Dibakar Datta Laboratory, New Jersey Institute of Technology},
  year         = {2026},
  url          = {https://dibakardatta.net/}
}

@misc{datta2026erroneous,
  author       = {Datta, Dibakar},
  title        = {No value is better than a wrong value: Detecting, repairing, and tolerating erroneous data in literature-mined datasets for {AI}-driven science},
  howpublished = {Preprint, Dibakar Datta Laboratory, New Jersey Institute of Technology},
  year         = {2026},
  url          = {https://dibakardatta.net/}
}

@misc{xie2005low,
  author        = {Xie, Daruo and Pfister, Olivier},
  title         = {Low-frequency measurements of electro-optic coefficients of {Na:KTP}, and {RTA}},
  year          = {2005},
  eprint        = {physics/0507123},
  archivePrefix = {arXiv},
  primaryClass  = {physics.optics},
  note          = {arXiv:physics/0507123v3}
}

@article{du2017electro,
  author  = {Du, Wan-Ying and Zhang, Pei and Zhang, Zi-Bo and Ren, Shuai and Wong, Wing-Han and Yu, Dao-Yin and Pun, Edwin Yue-Bun and Zhang, De-Long},
  title   = {Electro-optic coefficients $r_{13}$ and $r_{33}$ of singly {Er$^{3+}$}-doped and {In$^{3+}$/Er$^{3+}$}-codoped {LiNbO$_3$} crystals},
  journal = {Journal of Physics and Chemistry of Solids},
  volume  = {100},
  pages   = {101--106},
  year    = {2017},
  doi     = {10.1016/j.jpcs.2016.09.013}
}

@article{goulkov2003temperature,
  author  = {Goulkov, M. and Granzow, T. and D{\"o}rfler, U. and Woike, Th. and Imlau, M. and Pankrath, R. and Kleemann, W.},
  title   = {Temperature dependent determination of the linear electrooptic coefficient $r_{33}$ in {Sr$_{0.61}$Ba$_{0.39}$Nb$_2$O$_6$} single crystals by means of light-induced scattering},
  journal = {Optics Communications},
  volume  = {218},
  number  = {1--3},
  pages   = {173--182},
  year    = {2003},
  doi     = {10.1016/S0030-4018(03)01190-8}
}

@article{vantroeye2014first,
  author  = {Van Troeye, B. and Gillet, Y. and Ponc{\'e}, S. and Gonze, X.},
  title   = {First-principles characterization of the electronic and optical properties of hexagonal {LiIO$_3$}},
  journal = {Optical Materials},
  volume  = {36},
  number  = {9},
  pages   = {1494--1501},
  year    = {2014},
  doi     = {10.1016/j.optmat.2014.04.009}
}

@article{zhu2024first,
  author  = {Zhu, Jiachen and Liu, Tingyu and Hu, Hao and Zhao, Longfeng},
  title   = {First-principles study on the electronic structures and optical properties of {Mg}-doped {KDP} crystal},
  journal = {Materials Science in Semiconductor Processing},
  volume  = {170},
  pages   = {107972},
  year    = {2024},
  doi     = {10.1016/j.mssp.2023.107972}
}

@article{wen2024giant,
  author  = {Wen, Yiyang and Ren, Hongda and Du, Xiaona and Zhang, Yang},
  title   = {Giant electro-optic response in transparent rhombohedral ferroelectric {Sm-PIN-PMN-PT} crystal based on domain engineering},
  journal = {Journal of Applied Physics},
  volume  = {136},
  number  = {11},
  pages   = {113103},
  year    = {2024},
  doi     = {10.1063/5.0230598}
}

@article{gao2021electro,
  author  = {Gao, Yuan and Meshkat, Shadi and Johnston, Andrew and Zheng, Chao and Walters, Grant and Feng, Qixin and Wang, Xiaoping and Sun, Meng-Jia and Najarian, Amin Morteza and Xue, Dingjiang and Wang, Ya-Kun and Saidaminov, Makhsud I. and Voznyy, Oleksandr and Hoogland, Sjoerd and Sargent, Edward H.},
  title   = {Electro-Optic Modulation Using Metal-Free Perovskites},
  journal = {ACS Applied Materials \& Interfaces},
  volume  = {13},
  number  = {16},
  pages   = {19042--19047},
  year    = {2021},
  doi     = {10.1021/acsami.1c03406}
}

@article{abel2019large,
  author  = {Abel, Stefan and Eltes, Felix and Ortmann, J. Elliott and Messner, Andreas and Castera, Pau and Wagner, Tino and Urbonas, Darius and Rosa, Alvaro and Gutierrez, Ana M. and Tulli, Domenico and Ma, Ping and Baeuerle, Benedikt and Josten, Arne and Heni, Wolfgang and Caimi, Daniele and others},
  title   = {Large {Pockels} effect in micro- and nanostructured barium titanate integrated on silicon},
  journal = {Nature Materials},
  volume  = {18},
  number  = {1},
  pages   = {42--47},
  year    = {2019},
  doi     = {10.1038/s41563-018-0208-0}
}

@article{gao2019electro,
  author  = {Gao, Yuan and Walters, Grant and Qin, Ying and Chen, Bin and Min, Yimeng and Seifitokaldani, Ali and Sun, Bin and Todorovic, Petar and Saidaminov, Makhsud I. and Lough, Alan and Tongay, Sefaattin and Hoogland, Sjoerd and Sargent, Edward H.},
  title   = {Electro-Optic Modulation in Hybrid Metal Halide Perovskites},
  journal = {Advanced Materials},
  volume  = {31},
  number  = {16},
  pages   = {1808336},
  year    = {2019},
  doi     = {10.1002/adma.201808336}
}

@article{gao2021self,
  author  = {Gao, Yuan and Sachinthani, K. A. Niradha and Zheng, Chao and Jarrett-Wilkins, Charlie and Johnston, Andrew and Sun, Meng-Jia and Najarian, Amin Morteza and Wang, Ya-Kun and Saidaminov, Makhsud I. and Garc{\'i}a de Arquer, F. Pelayo and Seferos, Dwight S. and Hoogland, Sjoerd and Sargent, Edward H.},
  title   = {Self-Aligned Non-Centrosymmetric Conjugated Molecules Enable Electro-Optic Perovskites},
  journal = {Advanced Optical Materials},
  volume  = {9},
  number  = {20},
  pages   = {2100730},
  year    = {2021},
  doi     = {10.1002/adom.202100730}
}

@article{liu2024large,
  author  = {Liu, Xin and Li, Fangbing and Li, Jingwei and Tan, Peng and Lei, Jing and Zhao, Weigang and Song, Kexin and Zheng, Huaibin and Tian, Hao and Li, Fei and Wei, Xiaoyong and Xu, Zhuo},
  title   = {Large Electro-Optic Coefficient in Single-Domain {PIN-PMN-PT} Single Crystal},
  journal = {Advanced Optical Materials},
  volume  = {12},
  number  = {33},
  pages   = {2400885},
  year    = {2024},
  doi     = {10.1002/adom.202400885}
}

@article{wen2024enhanced,
  author  = {Wen, Yiyang and Cao, Yilin and Chen, Haisheng and Du, Xiaona and Guo, Jiaxing and Shen, Jiaying and Zhang, Fan and Bian, Taiyu and An, Yu and Ren, Hongda and Wu, Zhenping and Liu, Weiwei and Zhang, Yang},
  title   = {Enhanced Electro-Optic Coefficients in Single-Crystalline {BaTiO$_3$} Thin Films Enabled by Domain Alignment},
  journal = {Advanced Optical Materials},
  volume  = {12},
  number  = {15},
  pages   = {2303058},
  year    = {2024},
  doi     = {10.1002/adom.202303058}
}

@article{vakilitabatabaei2026electro,
  author  = {{Vakili Tabatabaei}, Mojdeh and Ojanen, Samu-Pekka and Viheri{\"a}l{\"a}, Jukka and Bhuiyan, Ifte Khairul Alam and Ruan, Yating and Komissinskiy, Philipp and Alff, Lambert and Guina, Mircea and Preu, Sascha},
  title   = {Electro-optically tunable barium--strontium--titanate filter for the 2~{$\mu$m} wavelength range},
  journal = {APL Photonics},
  volume  = {11},
  number  = {4},
  pages   = {046108},
  year    = {2026},
  doi     = {10.1063/5.0307690}
}

@article{zgonik1994dielectric,
  author  = {Zgonik, M. and Bernasconi, P. and Duelli, M. and Schlesser, R. and G{\"u}nter, P. and Garrett, M. H. and Rytz, D. and Zhu, Y. and Wu, X.},
  title   = {Dielectric, elastic, piezoelectric, electro-optic, and elasto-optic tensors of {BaTiO$_3$} crystals},
  journal = {Physical Review B},
  volume  = {50},
  number  = {9},
  pages   = {5941--5949},
  year    = {1994},
  doi     = {10.1103/PhysRevB.50.5941}
}

@article{cao2023characterization,
  author  = {Cao, Yu and Al Meselmene, Nour and Dogheche, Elhadj and Yang, Ping and Moitra, Parikshit and Li, Shi Qiang and Venkatesan, Thirumalai and Danner, Aaron},
  title   = {Characterization of electro-optic coefficients $r_{13}$, $r_{23}$ and $r_{33}$ in single crystalline {BaTiO$_3$} thin films},
  journal = {Optical Materials Express},
  volume  = {13},
  number  = {1},
  pages   = {152},
  year    = {2023},
  doi     = {10.1364/OME.474245}
}

@article{serafini2017photoconductive,
  author  = {Serafini, John and Hossain, A. and James, R. B. and Guziewicz, M. and Kruszka, R. and S{\l}ysz, W. and Kochanowska, D. and Domagala, J. Z. and Mycielski, A. and Sobolewski, Roman},
  title   = {Photoconductive and electro-optic effects in {(Cd,Mg)Te} single crystals measured in an experiment-on-chip configuration},
  journal = {Applied Physics Letters},
  volume  = {111},
  number  = {1},
  pages   = {011108},
  year    = {2017},
  doi     = {10.1063/1.4993155}
}

@article{kim2023nature,
  author  = {Kim, Inhwan and Paoletta, Therese and Demkov, Alexander A.},
  title   = {Nature of electro-optic response in tetragonal {BaTiO$_3$}},
  journal = {Physical Review B},
  volume  = {108},
  number  = {11},
  pages   = {115201},
  year    = {2023},
  doi     = {10.1103/PhysRevB.108.115201}
}

@article{zhu2016ferroelectric,
  author  = {Zhu, Minmin and Du, Zehui and Liu, Qing and Chen, Bensong and Tsang, Siu Hon and Teo, Edwin Hang Tong},
  title   = {Ferroelectric {BiFeO$_3$} thin-film optical modulators},
  journal = {Applied Physics Letters},
  volume  = {108},
  number  = {23},
  pages   = {233502},
  year    = {2016},
  doi     = {10.1063/1.4953201}
}

@article{shin2013ultralow,
  author  = {Shin, Jae Hyuk and Dagli, Nadir},
  title   = {Ultralow Drive Voltage Substrate Removed {GaAs/AlGaAs} Electro-Optic Modulators at 1550 nm},
  journal = {IEEE Journal of Selected Topics in Quantum Electronics},
  volume  = {19},
  number  = {6},
  pages   = {150--157},
  year    = {2013},
  doi     = {10.1109/JSTQE.2013.2263122}
}

@article{li2025engineering,
  author  = {Li, Haoyang and Thomas, Robert and Jiang, Pisu and Balram, Krishna C.},
  title   = {Engineering cm-scale true push-pull electro-optic modulators in a suspended {GaAs} photonic integrated circuit platform by exploiting the orientation induced asymmetry of the {Pockels} $r_{41}$ coefficient},
  journal = {Nanophotonics},
  volume  = {14},
  number  = {18},
  pages   = {3033--3042},
  year    = {2025},
  doi     = {10.1515/nanoph-2025-0212}
}

@inproceedings{krauser1986electro,
  author    = {Krauser, J. and Albrecht, P. and Bornholdt, C. and D{\"o}ldissen, W. and Niggebr{\"u}gge, U. and Nolting, H.-P. and Schlak, M.},
  title     = {Electro-optic modulators in {GaInAsP/InP}},
  booktitle = {Fiber Optic Sources and Detectors},
  series    = {Proc. SPIE},
  volume    = {587},
  pages     = {180},
  year      = {1986},
  doi       = {10.1117/12.951218}
}

@article{zgonik1993materials,
  author  = {Zgonik, M. and Schlesser, R. and Biaggio, I. and Voit, E. and Tscherry, J. and G{\"u}nter, P.},
  title   = {Materials constants of {KNbO$_3$} relevant for electro- and acousto-optics},
  journal = {Journal of Applied Physics},
  volume  = {74},
  number  = {2},
  pages   = {1287--1297},
  year    = {1993},
  doi     = {10.1063/1.354934}
}

@article{imai2007crystal,
  author  = {Imai, Tadayuki and Sasaura, Masahiro and Nakamura, Koichiro and Fujiura, Kazuo},
  title   = {Crystal Growth and Electro-optic Properties of {KTa$_{1-x}$Nb$_x$O$_3$}},
  journal = {NTT Technical Review},
  volume  = {5},
  number  = {9},
  pages   = {4--11},
  year    = {2007},
  doi     = {10.53829/ntr200709sp1}
}

@article{szabados2020electro,
  author  = {Szabados, Jan and Werner, Christoph S. and Herr, Simon J. and Breunig, Ingo and Buse, Karsten},
  title   = {Electro-optic eigenfrequency tuning of potassium tantalate-niobate microresonators},
  journal = {APL Photonics},
  volume  = {5},
  number  = {1},
  pages   = {016106},
  year    = {2020},
  doi     = {10.1063/1.5133029}
}

@article{zhou2023hydrothermal,
  author  = {Zhou, Haitao and He, Xiaoling and Wu, Wenyuan and Tong, Jingfang and Wang, Jinliang and Zuo, Yanbin and Wu, Yicheng and Zhang, Changlong and Hu, Zhanggui},
  title   = {Hydrothermal growth of {KTiOPO$_4$} crystal for electro-optical application},
  journal = {Light: Science \& Applications},
  volume  = {12},
  number  = {1},
  pages   = {23},
  year    = {2023},
  doi     = {10.1038/s41377-022-01022-0}
}

@article{jazbinsek2002material,
  author  = {Jazbin{\v{s}}ek, M. and Zgonik, M.},
  title   = {Material tensor parameters of {LiNbO$_3$} relevant for electro- and elasto-optics},
  journal = {Applied Physics B: Lasers and Optics},
  volume  = {74},
  number  = {4--5},
  pages   = {407--414},
  year    = {2002},
  doi     = {10.1007/s003400200818}
}

@article{lv2025new,
  author  = {Lv, Lingfei and Yang, Xiaoming and Wu, Hongjiang and Su, Rongbing and Wang, Zujian and Su, Bin and Shan, Pai and Li, Bingxuan and Long, Xifa and Zhang, Ge and Xu, Feng and He, Chao},
  title   = {A new-type electro-optic crystal: {K$_3$Nb$_3$B$_2$O$_{12}$}},
  journal = {Journal of Materiomics},
  volume  = {11},
  number  = {5},
  pages   = {100991},
  year    = {2025},
  doi     = {10.1016/j.jmat.2024.100991}
}

@article{majkic2017optical,
  author  = {Majki{\'c}, Aleksej and Franke, Alexander and Kirste, Ronny and Schlesser, Raoul and Collazo, Ram{\'o}n and Sitar, Zlatko and Zgonik, Marko},
  title   = {Optical nonlinear and electro-optical coefficients in bulk aluminium nitride single crystals},
  journal = {physica status solidi (b)},
  volume  = {254},
  number  = {9},
  pages   = {1700077},
  year    = {2017},
  doi     = {10.1002/pssb.201700077}
}

@article{goodno1995investigation,
  author  = {Goodno, G. D. and Guo, Z. and Miller, R. J. D. and Miller, I. J. and Montgomery, J. W. and Adhav, S. R. and Adhav, R. S.},
  title   = {Investigation of {$\beta$-BaB$_2$O$_4$} as a {Q} switch for high power applications},
  journal = {Applied Physics Letters},
  volume  = {66},
  number  = {13},
  pages   = {1575--1577},
  year    = {1995},
  doi     = {10.1063/1.113675}
}

@inproceedings{juvalta2007near,
  author    = {Juvalta, F. and Jazbin{\v{s}}ek, M. and G{\"u}nter, P. and Montemezzani, G. and Kitamura, K.},
  title     = {Near-stoichiometric {LiTaO$_3$} for deep {UV} electro-optical applications},
  booktitle = {2007 European Conference on Lasers and Electro-Optics and the International Quantum Electronics Conference},
  pages     = {1},
  year      = {2007},
  doi       = {10.1109/CLEOE-IQEC.2007.4386131}
}

@article{yezekyan2024plasmonic,
  author  = {Yezekyan, Torgom and Thomaschewski, Martin and Thrane, Paul Conrad Vaagen and Bozhevolnyi, Sergey I.},
  title   = {Plasmonic electro-optic modulators on lead zirconate titanate platform},
  journal = {Nanophotonics},
  volume  = {13},
  number  = {18},
  pages   = {3591--3598},
  year    = {2024},
  doi     = {10.1515/nanoph-2024-0039}
}

@misc{zhang2026ultrafast,
  author        = {Zhang, Yueyang and Li, Chenlei and Shu, Tao and Chen, Wei and Shi, Cunyu and Qiu, Feng and Dai, Daoxin},
  title         = {Ultrafast and long-term stability integrated {Pockels} laser with thin-film {PZT}},
  year          = {2026},
  eprint        = {2606.03450},
  archivePrefix = {arXiv},
  primaryClass  = {physics.optics},
  note          = {arXiv:2606.03450}
}

@article{paoletta2021pockels,
  author  = {Paoletta, Therese and Demkov, Alexander A.},
  title   = {{Pockels} effect in low-temperature rhombohedral {BaTiO$_3$}},
  journal = {Physical Review B},
  volume  = {103},
  number  = {1},
  pages   = {014303},
  year    = {2021},
  doi     = {10.1103/PhysRevB.103.014303}
}

@article{sando2014linear,
  author  = {Sando, D. and Hermet, P. and Allibe, J. and Bourderionnet, J. and Fusil, S. and Carr{\'e}t{\'e}ro, C. and Jacquet, E. and Mage, J.-C. and Dolfi, D. and Barth{\'e}l{\'e}my, A. and Ghosez, Ph. and Bibes, M.},
  title   = {Linear electro-optic effect in multiferroic {BiFeO$_3$} thin films},
  journal = {Physical Review B},
  volume  = {89},
  number  = {19},
  pages   = {195106},
  year    = {2014},
  doi     = {10.1103/PhysRevB.89.195106}
}

@incollection{cuniotponsard2011strontium,
  author    = {Cuniot-Ponsard, Mireille},
  title     = {Strontium Barium Niobate Thin Films for Dielectric and Electro-Optic Applications},
  booktitle = {Ferroelectrics -- Material Aspects},
  publisher = {InTech},
  isbn      = {978-953-307-332-3},
  year      = {2011},
  doi       = {10.5772/17321}
}

@article{powell2022integrated,
  author  = {Powell, Keith and Li, Liwei and Shams-Ansari, Amirhassan and Wang, Jianfu and Meng, Debin and Sinclair, Neil and Deng, Jiangdong and Lon{\v{c}}ar, Marko and Yi, Xiaoke},
  title   = {Integrated silicon carbide electro-optic modulator},
  journal = {Nature Communications},
  volume  = {13},
  number  = {1},
  pages   = {1851},
  year    = {2022},
  doi     = {10.1038/s41467-022-29448-5}
}

@article{ulrich2025engineering,
  author  = {Ulrich, Anja and Brahim, Kamal and Boelen, Andries and Debaets, Michiel and Khalil, Ahmed and Sun, Conglin and Huang, Yishu and Saseendran, Sandeep Seema and Baryshnikova, Marina and Favia, Paola and Nuytten, Thomas and Sergeant, Stefanie and Van Gasse, Kasper and Kuyken, Bart and De Greve, Kristiaan and others},
  title   = {Engineering high {Pockels} coefficients in thin-film strontium titanate for cryogenic quantum electro-optic applications},
  journal = {Science},
  volume  = {390},
  number  = {6771},
  pages   = {390--393},
  year    = {2025},
  doi     = {10.1126/science.adx3741}
}

@article{suceava2026colossal,
  author  = {Suceava, Albert and Hazra, Sankalpa and Ross, Aiden and Philippi, Ian Reed and Sotir, Dylan and Brower, Brynn and Ding, Lei and Zhu, Yingxin and Zhang, Zhiyu and Sarkar, Himirkanti and Sarker, Saugata and Yang, Yang and Sarker, Suchismita and Stoica, Vladimir A. and Schlom, Darrell G. and others},
  title   = {Colossal Cryogenic Electro-Optic Response Through Metastability in Strained {BaTiO$_3$} Thin Films},
  journal = {Advanced Materials},
  volume  = {38},
  number  = {3},
  pages   = {e07564},
  year    = {2026},
  doi     = {10.1002/adma.202507564}
}

@article{wang2022engineering,
  author  = {Wang, Sasa and Zhu, Tong and Sabatini, Randy and Najarian, Amin Morteza and Imran, Muhammad and Zhao, Ruyan and Xia, Pan and Zeng, Lewei and Hoogland, Sjoerd and Seferos, Dwight S. and Sargent, Edward H.},
  title   = {Engineering Electro-Optic {BaTiO$_3$} Nanocrystals via Efficient Doping},
  journal = {Advanced Materials},
  volume  = {34},
  number  = {47},
  pages   = {2207261},
  year    = {2022},
  doi     = {10.1002/adma.202207261}
}

@article{berozashvili1989dispersion,
  author  = {Berozashvili, Yu and Machavariani, S. and Natsvlishvili, A. and Chirakadze, A.},
  title   = {Dispersion of the linear electro-optic coefficients and the non-linear susceptibility in {GaP}},
  journal = {Journal of Physics D: Applied Physics},
  volume  = {22},
  number  = {5},
  pages   = {682--686},
  year    = {1989},
  doi     = {10.1088/0022-3727/22/5/017}
}

@article{zhu2018strong,
  author  = {Zhu, Minmin and Du, Zehui and Chng, Soon Siang and Tsang, Siu Hon and Teo, Edwin Hang Tong},
  title   = {Strong electro-optically active {Ni}-substituted {Pb(Zr$_{0.35}$Ti$_{0.65}$)O$_3$} thin films: toward integrated active and durable photonic devices},
  journal = {Journal of Materials Chemistry C},
  volume  = {6},
  number  = {47},
  pages   = {12919--12927},
  year    = {2018},
  doi     = {10.1039/C8TC04576C}
}

@article{jeon2007electro,
  author  = {Jeon, Yongseog and Kang, Hyun Shik},
  title   = {Electro-Optic Coefficient Measurements for {Zn$_x$Cd$_{1-x}$Te} Single Crystals at 1550 nm Wavelength},
  journal = {Optical Review},
  volume  = {14},
  number  = {6},
  pages   = {373--375},
  year    = {2007},
  doi     = {10.1007/s10043-007-0373-x}
}

@article{ion2023lead,
  author  = {Ion, Valentin and Teodorescu, Valentin and Birjega, Ruxandra and Dinescu, Maria and Mitterbauer, Christoph and Alexandrou, Ioannis and Ghitiu, Ioan and Craciun, Floriana and Scarisoreanu, Nicu D.},
  title   = {Lead-Free Perovskite Thin Films with Tailored {Pockels}--{Kerr} Effects for Photonics},
  journal = {ACS Applied Materials \& Interfaces},
  volume  = {15},
  number  = {31},
  pages   = {38039--38048},
  year    = {2023},
  doi     = {10.1021/acsami.3c06499}
}

@article{yang2024unveiling,
  author  = {Yang, Guangcanlan and Wang, Haochen and Mu, Sai and Xie, Hao and Wang, Tyler and He, Chengxing and Shen, Mohan and Liu, Mengxia and Van de Walle, Chris G. and Tang, Hong X.},
  title   = {Unveiling the {Pockels} coefficient of ferroelectric nitride {ScAlN}},
  journal = {Nature Communications},
  volume  = {15},
  number  = {1},
  pages   = {9538},
  year    = {2024},
  doi     = {10.1038/s41467-024-53895-x}
}

@article{hamze2020design,
  author  = {Hamze, Ali K. and Reynaud, Marc and Geler-Kremer, Jacqueline and Demkov, Alexander A.},
  title   = {Design rules for strong electro-optic materials},
  journal = {npj Computational Materials},
  volume  = {6},
  number  = {1},
  pages   = {130},
  year    = {2020},
  doi     = {10.1038/s41524-020-00399-z}
}

@article{chelladurai2025barium,
  author  = {Chelladurai, Daniel and Kohli, Manuel and Winiger, Joel and Moor, David and Messner, Andreas and Fedoryshyn, Yuriy and Eleraky, Mohammed and Liu, Yuqi and Wang, Hua and Leuthold, Juerg},
  title   = {Barium titanate and lithium niobate permittivity and {Pockels} coefficients from megahertz to sub-terahertz frequencies},
  journal = {Nature Materials},
  volume  = {24},
  number  = {6},
  pages   = {868--875},
  year    = {2025},
  doi     = {10.1038/s41563-025-02158-1}
}

@article{reynaud2025enhancement,
  author  = {Reynaud, Marc and Waqar, Moaz and Du, Chaojie and Posadas, Agham B. and Pan, Xiaoqing and Demkov, Alexander A.},
  title   = {Enhancement of electro-optic response in {BaTiO$_3$} films integrated on {Si} via heating and cooling rate control},
  journal = {Communications Materials},
  volume  = {6},
  number  = {1},
  pages   = {176},
  year    = {2025},
  doi     = {10.1038/s43246-025-00908-x}
}

@article{casson2004electro,
  author  = {Casson, Joanna L. and Gahagan, Kevin T. and Scrymgeour, David A. and Jain, Ravinder K. and Robinson, Jeanne M. and Gopalan, Venkatraman and Sander, Robert K.},
  title   = {Electro-optic coefficients of lithium tantalate at near-infrared wavelengths},
  journal = {Journal of the Optical Society of America B},
  volume  = {21},
  number  = {11},
  pages   = {1948},
  year    = {2004},
  doi     = {10.1364/JOSAB.21.001948}
}

@article{lauk2020perspectives,
  author  = {Lauk, Nikolai and Sinclair, Neil and Barzanjeh, Shabir and Covey, Jacob P. and Saffman, Mark and Spiropulu, Maria and Simon, Christoph},
  title   = {Perspectives on quantum transduction},
  journal = {Quantum Science and Technology},
  volume  = {5}, number = {2}, pages = {020501}, year = {2020},
  doi     = {10.1088/2058-9565/ab788a}}

@article{han2021microwave,
  author  = {Han, Xu and Fu, Wei and Zou, Chang-Ling and Jiang, Liang and Tang, Hong X.},
  title   = {Microwave-optical quantum frequency conversion},
  journal = {Optica}, volume = {8}, number = {8}, pages = {1050}, year = {2021},
  doi     = {10.1364/OPTICA.425414}}

@article{wang2018integrated,
  author  = {Wang, Cheng and Zhang, Mian and Chen, Xi and Bertrand, Maxime and Shams-Ansari, Amirhassan and Chandrasekhar, Sethumadhavan and Winzer, Peter and Lon{\v{c}}ar, Marko},
  title   = {Integrated lithium niobate electro-optic modulators operating at {CMOS}-compatible voltages},
  journal = {Nature}, volume = {562}, number = {7725}, pages = {101--104}, year = {2018},
  doi     = {10.1038/s41586-018-0551-y}}

@article{zhu2021integrated,
  author  = {Zhu, Di and Shao, Linbo and Yu, Mengjie and Cheng, Rebecca and Desiatov, Boris and Xin, C. J. and Hu, Yaowen and Holzgrafe, Jeffrey and Ghosh, Soumya and Shams-Ansari, Amirhassan and Puma, Eric and Sinclair, Neil and Reimer, Christian and Zhang, Mian and Lon{\v{c}}ar, Marko},
  title   = {Integrated photonics on thin-film lithium niobate},
  journal = {Advances in Optics and Photonics}, volume = {13}, number = {2}, pages = {242}, year = {2021},
  doi     = {10.1364/AOP.411024}}

@article{zhang2021integrated,
  author  = {Zhang, Mian and Wang, Cheng and Kharel, Prashanta and Zhu, Di and Lon{\v{c}}ar, Marko},
  title   = {Integrated lithium niobate electro-optic modulators: when performance meets scalability},
  journal = {Optica}, volume = {8}, number = {5}, pages = {652}, year = {2021},
  doi     = {10.1364/OPTICA.415762}}

@article{eltes2020integrated,
  author  = {Eltes, Felix and Villarreal-Garcia, Gerardo E. and Caimi, Daniele and Siegwart, Heinz and Gentile, Antonio A. and Hart, Andy and Stark, Pascal and Marshall, Graham D. and Thompson, Mark G. and Barreto, Jorge and Fompeyrine, Jean and Abel, Stefan},
  title   = {An integrated optical modulator operating at cryogenic temperatures},
  journal = {Nature Materials}, volume = {19}, number = {11}, pages = {1164--1168}, year = {2020},
  doi     = {10.1038/s41563-020-0725-5}}

@article{sahu2022quantum,
  author  = {Sahu, Rishabh and Hease, William and Rueda, Alfredo and Arnold, Georg and Qiu, Liu and Fink, Johannes M.},
  title   = {Quantum-enabled operation of a microwave-optical interface},
  journal = {Nature Communications}, volume = {13}, pages = {1276}, year = {2022},
  doi     = {10.1038/s41467-022-28924-2}}

@article{holzgrafe2020cavity,
  author  = {Holzgrafe, Jeffrey and Sinclair, Neil and Zhu, Di and Shams-Ansari, Amirhassan and Colangelo, Marco and Hu, Yaowen and Zhang, Mian and Berggren, Karl K. and Lon{\v{c}}ar, Marko},
  title   = {Cavity electro-optics in thin-film lithium niobate for efficient microwave-to-optical transduction},
  journal = {Optica}, volume = {7}, number = {12}, pages = {1714}, year = {2020},
  doi     = {10.1364/OPTICA.397513}}

@article{youssefi2021cryogenic,
  author  = {Youssefi, Amir and Shomroni, Itay and Joshi, Yash J. and Bernier, Nathan R. and Lukashchuk, Anton and Uhrich, Philipp and Qiu, Liu and Kippenberg, Tobias J.},
  title   = {A cryogenic electro-optic interconnect for superconducting devices},
  journal = {Nature Electronics}, volume = {4}, number = {5}, pages = {326--332}, year = {2021},
  doi     = {10.1038/s41928-021-00570-4}}

@article{dejong2015elastic,
  author  = {de Jong, Maarten and Chen, Wei and Angsten, Thomas and Jain, Anubhav and Notestine, Randy and Gamst, Anthony and Sluiter, Marcel and Ande, Chaitanya Krishna and van der Zwaag, Sybrand and Plata, Jose J. and Toher, Cormac and Curtarolo, Stefano and Ceder, Gerbrand and Persson, Kristin A. and Asta, Mark},
  title   = {Charting the complete elastic properties of inorganic crystalline compounds},
  journal = {Scientific Data}, volume = {2}, pages = {150009}, year = {2015},
  doi     = {10.1038/sdata.2015.9}}

@article{dejong2015piezo,
  author  = {de Jong, Maarten and Chen, Wei and Geerlings, Henry and Asta, Mark and Persson, Kristin Aslaug},
  title   = {A database to enable discovery and design of piezoelectric materials},
  journal = {Scientific Data}, volume = {2}, pages = {150053}, year = {2015},
  doi     = {10.1038/sdata.2015.53}}

@article{petousis2017high,
  author  = {Petousis, Ioannis and Mrdjenovich, David and Ballouz, Eric and Liu, Miao and Winston, Donald and Chen, Wei and Graf, Tanja and Schladt, Thomas D. and Persson, Kristin A. and Prinz, Fritz B.},
  title   = {High-throughput screening of inorganic compounds for the discovery of novel dielectric and optical materials},
  journal = {Scientific Data}, volume = {4}, pages = {160134}, year = {2017},
  doi     = {10.1038/sdata.2016.134}}

@article{basov2017towards,
  author  = {Basov, D. N. and Averitt, R. D. and Hsieh, D.},
  title   = {Towards properties on demand in quantum materials},
  journal = {Nature Materials}, volume = {16}, number = {11}, pages = {1077--1088}, year = {2017},
  doi     = {10.1038/nmat5017}}

@article{lu2000linear,
  author  = {Lu, Yu and Cheng, Zhong-Yang and Park, Seung-Eek and Liu, Shi-Fang and Zhang, Qiming},
  title   = {Linear electro-optic effect of 0.88{Pb(Zn$_{1/3}$Nb$_{2/3}$)O$_3$}--0.12{PbTiO$_3$} single crystal},
  journal = {Japanese Journal of Applied Physics}, volume = {39}, number = {1R}, pages = {141--145}, year = {2000},
  doi     = {10.1143/JJAP.39.141}}

@book{boyd2020nonlinear,
  author    = {Boyd, Robert W.},
  title     = {Nonlinear Optics},
  edition   = {4th},
  publisher = {Academic Press}, address = {London}, year = {2020}}

@book{yariv1984optical,
  author    = {Yariv, Amnon and Yeh, Pochi},
  title     = {Optical Waves in Crystals: Propagation and Control of Laser Radiation},
  publisher = {Wiley}, address = {New York}, year = {1984}}
